\documentclass[a4paper,fleqn,usenatbib]{mnras}

\usepackage{newtxtext,newtxmath}
\usepackage[T1]{fontenc}
\usepackage{ae,aecompl}

\usepackage{graphicx}	% Including figure files
\usepackage{amsmath}	% Advanced maths commands
\usepackage{amssymb}	% Extra maths symbols

\usepackage{calrsfs}

\title[Magnetized advective accretion flows]{ Magnetized advective accretion flows: formation of magnetic barriers in Magnetically Arrested Discs }

\author[T. Mondal and B. Mukhopadhyay]{
Tushar Mondal,$^{}$\thanks{E-mail: mtushar@iisc.ac.in}
Banibrata Mukhopadhyay$^{}$\thanks{E-mail: bm@iisc.ac.in}
\\
$^{}$Department of Physics, Indian Institute of Science, Bangalore 560012, India
}

\date{Accepted 2018 February 2. Received 2018 January 30; in original form 2017 October 20}

\pubyear{2018}

\begin{document}
\label{firstpage}
\pagerange{\pageref{firstpage}--\pageref{lastpage}}
\maketitle

% Abstract of the paper
\begin{abstract}

 We discuss the importance of large scale strong magnetic field in the removal of angular momentum outward, as well as the possible origin of different kinds of magnetic barrier in advective, geometrically thick, sub-Keplerian accretion flows around black holes. The origin of this large scale strong magnetic field near the event horizon is due to the advection of the magnetic flux by the accreting gas from the environment, say, the interstellar medium or a companion star, because of flux freezing. In this simplest vertically averaged, $1.5-$dimensional disc model, we choose the maximum upper limit of the magnetic field, which the disc around a black hole can sustain. In this so called magnetically arrested disc (MAD) model, the accreting gas either decelerates or faces the magnetic barrier near the event horizon by the accumulated magnetic field depending on the geometry. The magnetic barrier may knock the matter to infinity. We suggest that these types of flow are the building block to produce jets and outflows in the accreting system. We also find that in some cases, when matter is trying to go back to infinity after knocking the barrier, matter is prevented being escaped by the cumulative action of strong gravity and the magnetic tension, hence by another barrier. In this way, magnetic field can lock the matter in between these two barriers and it might be a possible explanation for the formation of episodic jet.

\end{abstract}

% Select between one and six entries from the list of approved keywords.
% Don't make up new ones.
\begin{keywords}
accretion, accretion discs -- black hole physics -- MHD (magnetohydrodynamics)
\end{keywords}

%%%%%%%%%%%%%%%%%%%%%%%%%%%%%%%%%%%%%%%%%%%%%%%%%%

%%%%%%%%%%%%%%%%% BODY OF PAPER %%%%%%%%%%%%%%%%%%

\section{Introduction}

\cite{Bekenstein1972}, based on the suggestion given by Geroch in 1971, argued
for an engine, namely Geroch-Bekenstein engine, converting mass to energy with 
an almost $100\%$ efficiency in the extreme gravitational potential of a black hole. 
While such an efficient conversion is generally very difficult in accretion 
disc, \cite{Bisnovatyi-Kogan1974,Bisnovatyi-Kogan1976} and later \cite{Narayan2003} by numerical simulations showed that it is possible in the presence
of very strong large-scale fields. Such a strongly
magnetic field dominated flow with a significant advection drags poloidal
magnetic fields to the inner region owing to flux freezing. This is expected to
result in the accumulation of significant field tending to disrupt the axisymmetric 
accretion flow relatively far away from the black hole. Inside the radius 
of disruption the matter is shown to accrete as discrete blobs with a velocity 
much smaller than the free-fall velocity and almost the entire rest mass 
of infalling matter is converted to energy, similar to Geroch-Bekenstein engine. 
Latter authors, mentioned above, named such an accretion flow as Magnetically Arrested Disc (MAD).
Note that radiatively efficient Keplerian discs are cooler and,
hence, the magnetic field would slip through the matter by means of ambipolar 
diffusion preventing the accumulation of fields. The same is true if anomalous 
magnetic diffusivity is large \citep{Lovelace1994,Lubow1994}. 

Now relativistic jets are very common in accreting black hole sources,
observed in stellar-mass black hole (for a review, see \citealt{Remillard2006}) 
as well as supermassive black hole in particular AGN \citep{Tremaine2002} sources.
Sometimes powerful jets are observed with energy more than the Eddington limit of the 
black hole \citep{Rawlings1991, Ghisellini2010, McNamara2011}, which argues for an efficient engine 
lying with their formation. Indeed the signature of dynamically important magnetic
fields was found in the black hole source at the center of our galaxy 
\citep{Eatough2013} and also based on the correlation of jet magnetic fields with
accretion disc luminosity for 76 radio-loud galaxies, the jet
launching region was concluded to exhibit dynamically important magnetic fields \citep{Zamaninasab2014}. All these observations/inferences support the idea of MAD. 

By global three-dimensional non-radiative 
general relativistic magnetohydrodynamics (GRMHD) simulations, \cite{Tchekhovskoy2011} described MAD impeding accretion and producing efficient
outflowing energy in jets and winds. They showed the combined effects of large-scale 
magnetic fields and spin of black hole could produce energy even more than $100\%$.
Of course crossing the outflowing energy beyond what is supplied is due to the fast
spinning nature of black hole. However, the spin effect of black hole \citep{Blandford1977} is required to be supplemented by magnetic field to reveal
higher and higher efficiency which is possible for underlying MAD.
The ordered magnetic field threading with disc matter corotates with accreting
material and also helps to power the jets via gravitational energy released \citep{Blandford1982}. However, there is also possible outer movement of magnetic fields
\citep{Bisnovatyi-Kogan1974, Bisnovatyi-Kogan1976, van Ballegooijen1989, Lubow1994, Ogilvie2001} and final inner accumulation of field lines is determined
by the balance between advection and outward diffusion of large-scale magnetic fields.
Nevertheless, it was shown that significant inward dragging of fields is possible
in flows having significant advection, unlike Keplerian flows (see also \citealt{Cao2011}).

When flow acquires strong magnetic fields, in particular far away from the
black hole, magnetic tension and hence the corresponding shearing stress 
could be significant. Hence, the underlying Maxwell stress could play role
in transporting angular momentum outward and matter inward, as viscous
shearing stress does (in the presence of molecular viscosity). \cite{Mukhopadhyay2015} showed that large-scale magnetic fields indeed can transport angular momentum
in advective accretion flows as efficiently as the $\alpha$-viscosity \citep{Shakura1973} does. Nevertheless, their choice of field strength was
low enough in particular the inner region of flow to have weaker magnetic barrier
revealing MAD. As a result there is continuous accretion in that model framework
due to large-scale magnetic fields. 

 In the present paper, we plan to explore large-scale stronger magnetic fields
to transport angular momentum as well as possibility of magnetic barrier created 
eventually in the inner flow region. 
Here, Alfv\'enic critical points control the flow behaviour, rather than 
fast magnetosonic points as was for the weaker field case discussed by \cite{Mukhopadhyay2015}.

 The plan of this paper is as follows. In \S \ref{sec:equations} we present the basic equations for magnetized advective accretion flow considering magnetic heating term in more general way. In \S \ref{sec:sonic} we apply those equations to evaluate the critical point conditions. In \S \ref{sec:result} we discuss the results including both the disc flow behaviours and the origin of different magnetic barriers. In \S \ref{sec:discussion} we summarize and give overall conclusions.

\section{HEIGHT-AVERAGED EQUATIONS OF MAGNETIZED ADVECTIVE ACCRETION FLOWS}\label{sec:equations}

Following standard practice, we vertically average the geometrically thick accretion flow equations and consider the motion to be confined in the two-dimensional equatorial $r-\phi$ plane. We assume a steady and axisymmetric flow such that $\partial/\partial t\equiv \partial/\partial \phi \equiv 0$ and all the dynamical flow parameters, namely, radial velocity ($v$), specific angular momentum ($\lambda$), mass density ($\rho$), fluid pressure ($p$), radial ($B_{r}$) and azimuthal ($B_{\phi}$) components of magnetic field are functions of radial coordinate $r$ only. 

Here, we plan to investigate the effects of large scale strong magnetic fields on the advective accretion flows in order to transport matter, as well as the possible origin of magnetic barrier supporting ``magnetically arrested disc" (MAD) model, in the pseudo-Newtonian framework with \citet{Mukhopadhyay2002} potential. The choice of this potential allows us to use the Newtonian framework, whereas capturing certain important features of general relativity quite accurately, compared to that would appear in the full general relativistic framework.

Throughout in our calculation, we express the radial and vertical coordinates in units of $GM_{BH}/c^{2}$, where $G$ is the Newton's gravitational constant, $M_{BH}$ is the mass of the black hole and $c$ is the speed of light. We also express the velocity in units of $c$ and the specific angular momentum in $GM_{BH}/c$ to make all the variables dimensionless. Hence, the equation of continuity, the radial and azimuthal components of momentum equation and the energy equation are, respectively,
\begin{eqnarray}
&\frac{d}{dr}\left(r\rho h v\right)=0,\label{eq:continuity}& \\ &v\frac{dv}{dr}-\frac{\lambda^{2}}{r^{3}}+\frac{1}{\rho}\frac{dp}{dr}+F=-\frac{B_{\phi}}{4 \pi \rho}\left(\frac{dB_{\phi}}{dr}+\frac{B_{\phi}}{r}\right),\label{eq:radmomentum}& \\ &v\frac{d\lambda}{dr}=\frac{1}{r\rho h}\frac{d}{dr}\left(r^{2}W_{r\phi}h\right)+\frac{B_{r}}{4 \pi \rho h}\frac{d}{dr}\left(r B_{\phi} h\right),\label{eq:angmomentum}& \\ &\Sigma vT\frac{ds}{dr}=\frac{h(r)v}{\Gamma_{3}-1}\left(\frac{dp}{dr}-\frac{\Gamma_{1} p}{\rho}\frac{d\rho}{dr}\right)=Q^{+}-Q^{-}=f_{m}Q^{+}. \label{eq:energy}&
\end{eqnarray}
Here, $F$ is the magnitude of gravitational pseudo-Newtonian force given by \citet{Mukhopadhyay2002} as 
\begin{eqnarray*}
F=\frac{(r^{2}-2a\sqrt{r}+a^{2})^{2}}{r^{3}\lbrace \sqrt{r}(r-2)+a\rbrace ^{2}}\,,
\end{eqnarray*}
where throughout in our calculation the Kerr-parameter $a=0$ as for the non-rotating black hole (same as \citealt{Paczy1980}), $W_{r\phi}$ is the viscous shearing stress which can be written using \citet{Shakura1973} prescription with appropriate modification given by  \citet{MukhopadhyayGhosh2003}, and from vertical equilibrium assumption, the half-thickness of the disc in the presence of magnetic field can be written as
\begin{eqnarray}
h(r)= r^{1/2} F^{-1/2} \sqrt{\left(p+\frac{B^{2}}{8\pi}\right)\big/\rho}.
\end{eqnarray} 
In equation~(\ref{eq:energy}), the left-hand side is the radial advected entropy, where $\Sigma$ is the vertically integrated mass density, $T$ is the (ion) temperature and $s$ is the entropy density of the flow. Here, the adiabatic exponents can be written as \citep[e.g.][]{Chandrasekhar1939} 
\begin{eqnarray*}
\Gamma_{1}=\beta + \frac{\left(4-3\beta\right)^{2}\left(\gamma-1\right)}{\beta+12\left(\gamma-1\right)\left(1-\beta\right)},\qquad \Gamma_{3}=\frac{\Gamma_{1}-\beta}{4-3\beta},
\end{eqnarray*}
where $\gamma$ is the ratio of the specific heats and $\beta$ is the ratio of gas pressure to total pressure, given by
\begin{eqnarray*}
\beta=\frac{\rho k_{B}T/\mu m_{p}}{\overline{a}T^{4}/3+\rho k_{B}T/\mu m_{p}+B^{2}/8\pi}.
\end{eqnarray*}
Here, $\overline{a}$ is the Stefan constant, $k_{B}$ is the Boltzmann constant, $\mu$ is the mean molecular weight and $m_{p}$ is the mass of proton. In the two limiting cases, for a gas pressure-dominated flow $\beta=1$ and $\Gamma_{1}=\gamma=\Gamma_{3}$, and for a radiation-dominated flow $\beta=0$ and $\Gamma_{1}=4/3=\Gamma_{3}$. The right hand side of equation~(\ref{eq:energy}) gives the difference between the net rate of heat energy generated per unit area $Q^{+}$ and the energy radiated out per unit area $Q^{-}$, while the energy generated term can be written as $Q^{+}=Q^{+}_{vis}+Q^{+}_{mag},$ whereas the contribution comes from both viscous and magnetic effects. The details about viscous contribution are given in the existing literature \citep[e.g.][]{Chakrabarti1996,Mukhopadhyay2015}. An abundant supply of magnetic energy and the annihilation of the magnetic fields are responsible for magnetic heating and this magnetic heating contribution per unit area is given by \citep[e.g.][]{Bisnovatyi-Kogan1974,Choudhuri1998,BalbusHawley1998,Mukhopadhyay2015}
\begin{eqnarray}
Q^{+}_{mag}=\frac{h(r)}{4 \pi}\left[B_{r}^{2} \frac{dv}{dr}+B_{\phi} B_{r} \left(\frac{1}{r}\frac{d\lambda}{dr}-\frac{\lambda}{r^{2}}\right)\right].
\end{eqnarray}
The factor $f_{m}$ measures the degree to which the flow is cooling-dominated or advection-dominated (see \citealt{Narayan1994}) and it varies from $0$ to $1$. In the extreme limit of very efficient cooling, $f_{m}=0$, while for no cooling $f_{m}=1.$ 

In order to have a full dynamical theory of magnetohydrodynamics (MHD) flows, we now require two more equations, namely, magnetic induction equation and equation for no magnetic monopole, given by, respectively 
\begin{eqnarray}
&\nabla \times \left(\mathbf{v}\times\mathbf{B} \right)+\nu_{m}\nabla^{2}\mathbf{B}=0, \label{eq:induction}& \\ &\frac{d}{dr}\left(r B_{r}\right)=0, \label{eq:nomonopole}&
\end{eqnarray}
where $\mathbf{v}$ and $\mathbf{B}$ are respectively velocity and magnetic field vectors and $\nu_{m}$ is the magnetic diffusivity. On taking the ratio of the orders of first to second terms of the left hand side in equation~(\ref{eq:induction}), we obtain the dimensionless number $\mathcal{R}_{m}=LV/\nu_{m},$ known as magnetic Reynolds number, when $L$ and $V$ are respectively the characteristic length scale and velocity of the system. When, $\mathcal{R}_{m}$ is very large, which is the case for accretion disc, the second term of equation~(\ref{eq:induction}) can be neglected. Hence, the induction equation becomes
\begin{eqnarray}
\frac{d}{dr}\left(vB_{\phi}-\frac{\lambda B_{r}}{r}\right)=0. \label{eq:induction2}
\end{eqnarray}
To obtain the full dynamical solutions, the initial and the boundary conditions are very important. For the present purpose at the beginning of the sub-Keplerian flow, far away from the black hole at the transition radius, $\lambda=\lambda_{K}$ (where $\lambda_{K}$ being the Keplerian angular momentum per unit mass) which corresponds to the outer boundary $r=r_{out}$, whereas the event horizon of the black hole is the inner boundary, where the velocity becomes of the order of unity. In addition, an important condition has to be supplied at the magnetosonic radius discussed in \S \ref{sec:sonic}. We also have to supply $M_{BH}$, $\dot{M}$, $f_{m}$, $\alpha$ and $\gamma$ for a flow, where $\dot{M}$ is the constant mass accretion rate and $\alpha$ is the \cite{Shakura1973} viscosity parameter. 

\section{SOLUTION PROCEDURE AND MAGNETOSONIC POINT ANALYSIS}\label{sec:sonic}

The set of six coupled differential equations $(1)$, $(2)$, $(3)$, $(4)$, $(8)$ and $(9)$ can be solved simultaneously using appropriate boundary conditions including that at sonic/critical point(s) to obtain the solutions for six important dynamical variables: $v$, $\lambda$, $B_{r}$, $B_{\phi}$, $p$ and $\rho$, as functions of the independent variable $r$. Note that in the presence of strong magnetic fields
giving rise to magnetic shearing stress considered here, magnetorotational instability is expected to be suppressed and hence 
$\alpha\sim 0$. On the other hand, it can also be checked that for a
reasonable value of $\alpha$, the second (magnetic) term in the right hand side of equation (\ref{eq:angmomentum}) is generally 
at least an order of magnitude higher than the 
first (nonmagnetic viscous) term for the fields eventually considered in the subsequent sections. Thus in the rest of the computation we assume $\alpha=0$. 
Indeed our main aim here is to examine the flow behaviour,
underlying possible transport etc., solely via strong magnetic fields. Now, combining all the above equations, we can write $dv/dr$ in terms of other dynamical variables and the independent variable $r$ only, such as 
\begin{eqnarray}
\frac{dv}{dr}=\frac{\mathcal{N}}{\mathcal{D}}=\frac{\mathcal{N}}{\mathcal{A}v^{4}+\mathcal{B}v^{2}+\mathcal{C}}. \label{eq:slope}
\end{eqnarray}
where the numerator $\mathcal{N}$ is
\begin{eqnarray*}
&\mathcal{N}=\frac{p}{\rho}\left(2\frac{p}{\rho}+v_{A}^{2}\right)\frac{1}{F}\frac{dF}{dr}v\left(v_{Ar}^{2}-v^{2}\right)\Gamma_{1}+ \\ 
&Fv\left(v_{Ar}^{2}-v^{2}\right)\left(2\frac{p}{\rho}(1+\Gamma_{1})+v_{A}^{2}\right)+ \\ 
&2v^{2}v_{Ar}v_{A\phi}\left(2\frac{p}{\rho}+v_{A}^{2}\right)\frac{\lambda}{r^{2}}+ \\ 
&\left(2\frac{p}{\rho}+v_{A}^{2}\right)\frac{\lambda^{2}}{r^{3}}v\left(v^{2}-v_{Ar}^{2}\right)+ \\
&\frac{p}{\rho}\left(\frac{6p/\rho}{r}+\frac{v_{A}^{2}}{r}+2\frac{\lambda^{2}}{r^{3}}\right)\Gamma_{1}v\left(v^{2}-v_{Ar}^{2}\right)- \\
&\left(2\frac{p}{\rho}+v_{A}^{2}\right)v^{3}v_{A\phi}^{2}/r+ \\
&\left(2\frac{p}{\rho}+v_{A}^{2}\right)f_{m}\left(v^{2}\lambda +v_{Ar}^{2}\lambda - rv_{Ar}v_{A\phi}v \right)\left(\Gamma_{3}-1\right)\frac{v_{Ar}v_{A\phi}}{r^{2}},
\end{eqnarray*}
when $v_{Ar}=B_{r}/\sqrt{4\pi \rho}$, $v_{A\phi}=B_{\phi}/\sqrt{4\pi \rho}$ and the Alfv\'en velocity $v_{A}=\sqrt{v_{Ar}^{2}+v_{A\phi}^{2}}.$ The coefficients of the denominator $\mathcal{D}$ are
\begin{eqnarray*}
&\mathcal{A}=\left(2\frac{p}{\rho}(1+\Gamma_{1})+v_{A}^{2}\right), \\
&\mathcal{B}=\left(2\frac{p}{\rho}+v_{A}^{2}\right)\left(v_{Ar}^{2}f_{m}\left(\Gamma_{3}-1\right)-v_{A}^{2}-2\frac{p}{\rho}\Gamma_{1}\right)-2v_{Ar}^{2}\frac{p}{\rho}\Gamma_{1}, \\
&\mathcal{C}=v_{Ar}^{2}\left(2\frac{p}{\rho}+v_{A}^{2}\right)\left(2\frac{p}{\rho}\Gamma_{1}-f_{m}v_{A}^{2}\left(\Gamma_{3}-1\right)\right).
\end{eqnarray*}

To guarantee a smooth solution around a point where $\mathcal{D}=0$, $\mathcal{N}$ must be vanished therein. These points are called ``critical points", where $r=r_{c}$. Also the variables with subscript `$c$' refer to the values of that respective variables at that critical radius. Since at $r=r_{c}$, $dv/dr=0/0$, using l'Hospital's rule and after some algebra, it is easy to show that the velocity gradient at the critical point $(dv/dr)_{c}$ of the accreting matter has two values: one is valid for accretion solution and other for wind. The nature of the critical point depends on the values of the velocity gradient at the critical point. When both the velocity gradients are real and of opposite sign, the critical point is `saddle' type. When the gradients are real and of same sign, the critical point is `nodal' type. When the the gradients are complex, the critical point is `spiral' type (or `O'-type for non-dissipative system). For details of the classifications, see \citealt{Chakrabarti1990}.

We assume, the Alfv\'en velocity at $r_{c}$ to be expressed in terms of sound speed $c_{sc}$ as
\begin{eqnarray}
v_{Arc}=\frac{c_{sc}}{f_{r}\sqrt{2}},\\
v_{A\phi c}=\frac{c_{sc}}{f_{\phi}\sqrt{2}},
\end{eqnarray}
where $c_{s} \simeq \sqrt{\frac{p}{\rho}}$, the factor $\sqrt{2}$ is the normalization factor and the constants $f_{r}$ and $f_{\phi}$ imply the inverse of the magnetic field strength.
 
\begin{figure*}
\includegraphics[width=16cm]{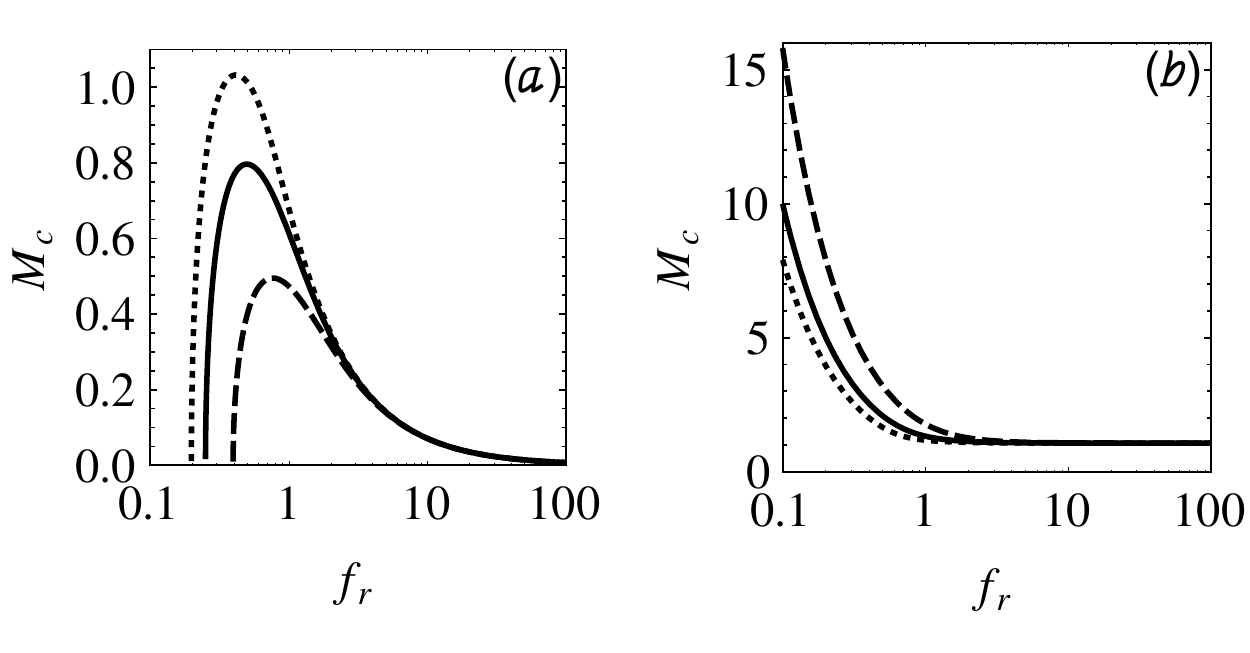}\caption{The variation of Mach number  at critical point as a function of magnetic field strength. The constant factor $f_{r}$ implies inverse of the field strength, $(a)$ for Alfv\'en wave, $(b)$ for fast magnetosonic wave, when the different lines are for different relative strength of magnetic fields (radial and azimuthal) at the critical point such as $B_{rc}=B_{\phi c}$ (solid lines), $B_{rc}=B_{\phi c}/2$ (dashed lines) and $B_{rc}=2B_{\phi c}$ (dotted lines). The other parameters are $M_{BH}=10M_{\odot}, \ \dot{M}=0.01\dot{M}_{Edd}$ and  $f_{m}=0.5$.} 
    \label{fig:figure1}
\end{figure*}

In general, the steady MHD flow reveals three critical points, at which the radial velocity is of the order of the propagation speed of each of the three different types of mode - the fast magnetosonic wave, the Alfv\'en wave and the slow magnetosonic wave { (\citealt{Weber1967}; also see \citealt{sakurai,Das2007})}. The Alfv\'en wave is purely transverse and the magnetic tension is the only restoring force for it. The other two magnetosonic waves are the mixtures of acoustic and magnetic waves. The slow magnetosonic point is absent here due the cold nature of the flow \citep[e.g.][]{Li1992,Gammie1999}. From $\mathcal{D}=0$, we can obtain the expressions for Mach number $(M_{c})$, the ratio of radial velocity $(v)$ to sound speed $(c_{s})$, at the critical point for two different modes: Alfv\'enic and fast magnetosonic. Figure~\ref{fig:figure1} shows the variation of $M_{c}$ with the change of the constant parameter $f_{r}$, for different relative strengths of radial and azimuthal components of the magnetic field obtained by adjusting the other constant parameter $f_{\phi}$. Figure~\ref{fig:figure1}$(a)$ is for the Alfv\'enic mode, whereas Figure~\ref{fig:figure1}$(b)$ is for the fast magnetosonic mode. Figure~\ref{fig:figure1}$(a)$ signifies that $M_{c}$ corresponding to the Alfv\'enic mode shrinks to disappear when $f_{r}$ is very large (which corresponds to the week magnetic field). On the other hand in Figure~\ref{fig:figure1}$(b)$, in this large $f_{r}$ limit, $M_{c}$ corresponding to the fast magnetosonic mode becomes unity and hence the disc behaves like of a simple hydrodynamics type. Now, in the lower $f_{r}$ limit (corresponding to the very strong magnetic field), for Alfv\'enic mode, $v$ is a more sensible parameter compared to $c_{s}$ depending on either the disc is poloidally or toroidally dominated. For the $B_{r}$ dominated case (dotted line), matter drags inward more rather than orbital circulation, on the other hand, for the $B_{\phi}$ dominated case (dashed line), matter rotates more rather than its inward acceleration making $v$ less. Hence, $M_{c}$ is higher for $B_{r}$ dominated case (dotted line) compared to $B_{\phi}$ dominated case (dashed line). However, in Figure~\ref{fig:figure1}$(b)$, for the fast magnetosonic mode, $v$ as well as $c_{s}$ both are sensible parameters for different relative strengths. Here the matter density is very high in the $B_{r}$ dominated case (dotted line) compared to the $B_{\phi}$ dominated case (dashed line) making $c_{s}$ as well as $v$ smaller for former. For the Alfv\'enic mode, the critical Mach number profile shows a maximum and hence it decreases with lowering the value of $f_{r}$. This is because of the absence of the vertical magnetic field. Since we consider purely the disc (which is vertically integrated), the disc sustains all the magnetic field lines in the two-dimensional flow only, unlike what could be in the presence of the vertical motion. \cite{Mukhopadhyay2015} already initiated exploring the disc dynamics for large $f_{r}$ $(\sim 100)$ and hence for the weak field limit, of fast magnetosonic mode. Here, we plan to address the dynamics for small $f_{r}$, mostly around less than unity, for the Alfv\'enic mode.

\section{RESULTS}\label{sec:result}

\citet{Mukhopadhyay2015} showed that the removal of angular momentum is possible in the presence of large scale magnetic stress in geometrically thick, advective, sub-Keplerian accretion flow, in the complete absence of $\alpha$-viscosity. It was suggested that the externally generated large-scale poloidal magnetic field, originating from the environment, say, the interstellar medium, would be dragged inward and greatly squeezed near the black hole by the accreting plasma \citep[e.g.][]{Bisnovatyi-Kogan1974,Bisnovatyi-Kogan1976}. In this case, when the large scale magnetic field is strong enough, the accretion flow will be arrested by the magnetic field in the inner region of the disc and it modifies the disc structure such that it becomes { a MAD \citep[e.g.][]{Narayan2003,Igumenshchev2008,Mck12}.}
 
Here, we plan to understand the followings. $(1)$ What is the nature of the accretion flow near the black hole in the presence of large scale strong magnetic field. $(2)$ Will there be any magnetic barrier, such that accretion will stop? $(3)$ What will be the fate of matter after knocking the barrier? Will it again go back to infinity?

\begin{figure*}
\includegraphics[width=16cm]{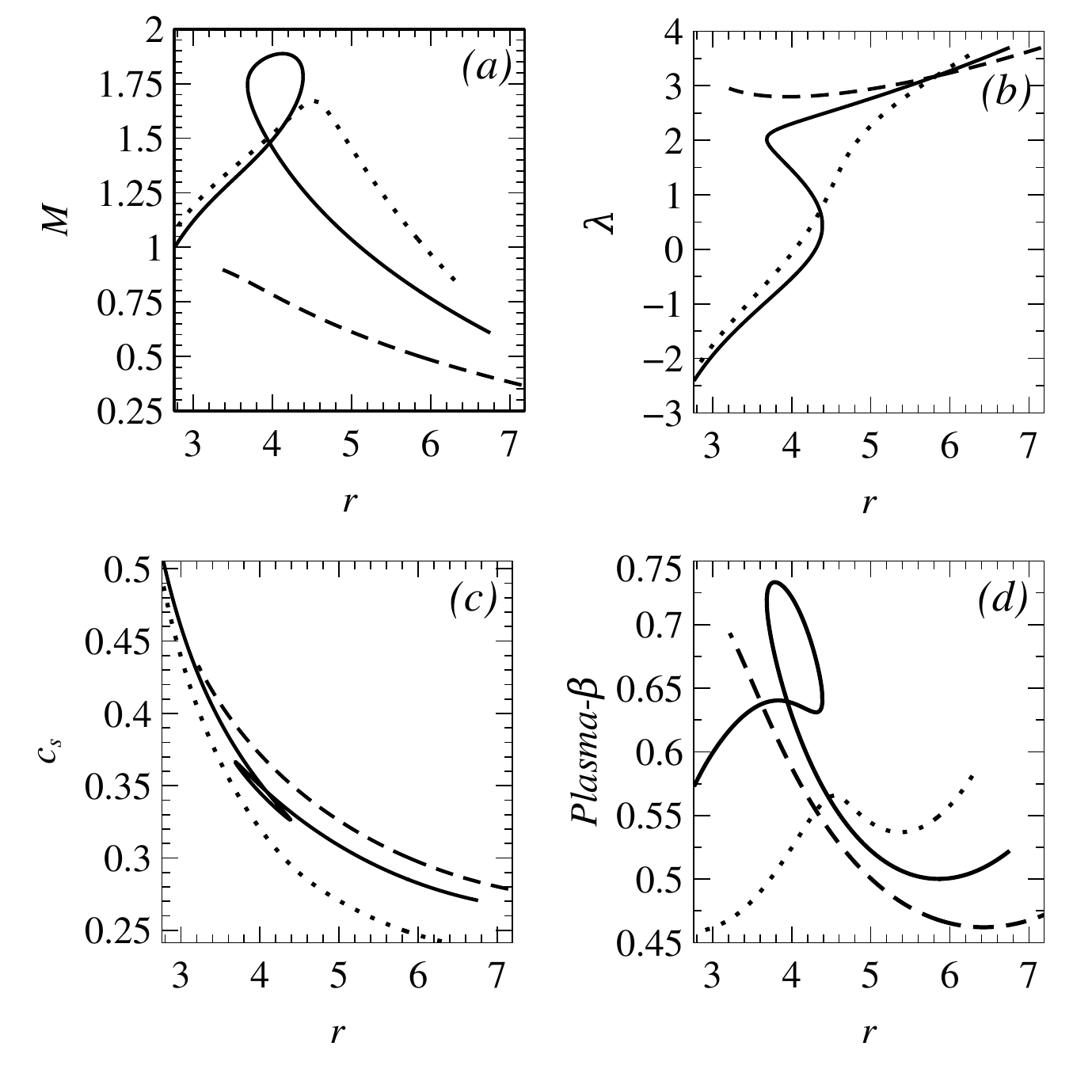}\caption{$(a)$ Mach number, $(b)$ angular momentum per unit mass, $(c)$ sound speed and $(d)$ plasma-$\beta$, when the different lines are for different relative strength of magnetic field (radial and azimuthal) at the critical point such as $B_{rc}=B_{\phi c}$ (solid lines), $B_{rc}=B_{\phi c}/2$ (dashed lines) and $B_{rc}=2B_{\phi c}$ (dotted lines). Here, $\lambda_{c}=3.2$. The other parameters are $M_{BH}=10M_{\odot}, \ \dot{M}=0.01\dot{M}_{Edd}$ and $f_{m}=0.5$.} 
    \label{fig:figure2}
\end{figure*}

\subsection{The origin of magnetic barrier}\label{sec:barrier}

To get the idea of different types of magnetic barrier and their origin, we have to understand the contribution from all the forces, say, gravitational force in the pseudo-Newtonian regime, the centrifugal force, forces due to fluid pressure and from magnetic fields, on the accretion phenomenon in the presence of large scale magnetic field. For this purpose, we have to look at the radial momentum balance equation~(\ref{eq:radmomentum}) more carefully. The first term on R.H.S. of this equation comes from magnetic pressure, whereas the second term from magnetic tension. Now, it is very easy to understand that the magnetic tension will always support gravity. What about the magnetic pressure, which generally acts against gravity, as like normal fluid pressure (see \citealt{Spruit2013})? The terms associated with radial magnetic field from pressure and tension parts are exactly equal and opposite, hence they cancel each other. In this circumstances, the profile for azimuthal component plays the key role to create any kind of magnetic barrier. The accretion process suppresses, when the forces against gravity dominate, such that
\begin{eqnarray}
-\frac{1}{4\pi \rho}\left(B_{\phi}\frac{dB_{\phi}}{dr}\right)-\frac{1}{\rho}\frac{dp}{dr}+\frac{\lambda^{2}}{r^{3}}>F+\frac{1}{4\pi \rho}\frac{B_{\phi}^{2}}{r}.
\label{eq:barrierIaI}
\end{eqnarray}
Hence, the essential conditions for which matter facing the barrier are
\begin{eqnarray}
B_{\phi}\frac{dB_{\phi}}{dr}<0 \quad \textrm{and} \quad \Bigg\vert B_{\phi}\frac{dB_{\phi}}{dr}\Bigg\vert \gg \frac{B_{\phi}^{2}}{r} \label{eq:barrierIa}.
\end{eqnarray}
Combining these, we obtain $\frac{dB_{\phi}}{dr} \ll -\frac{B_{\phi}}{r}$ and from equation~(\ref{eq:nomonopole}) we already know, $\frac{dB_{r}}{dr} = -\frac{B_{r}}{r}.$ Hence, the barrier appears, when the disc is poloidally dominated. 

\begin{figure*}
\includegraphics[width=16cm]{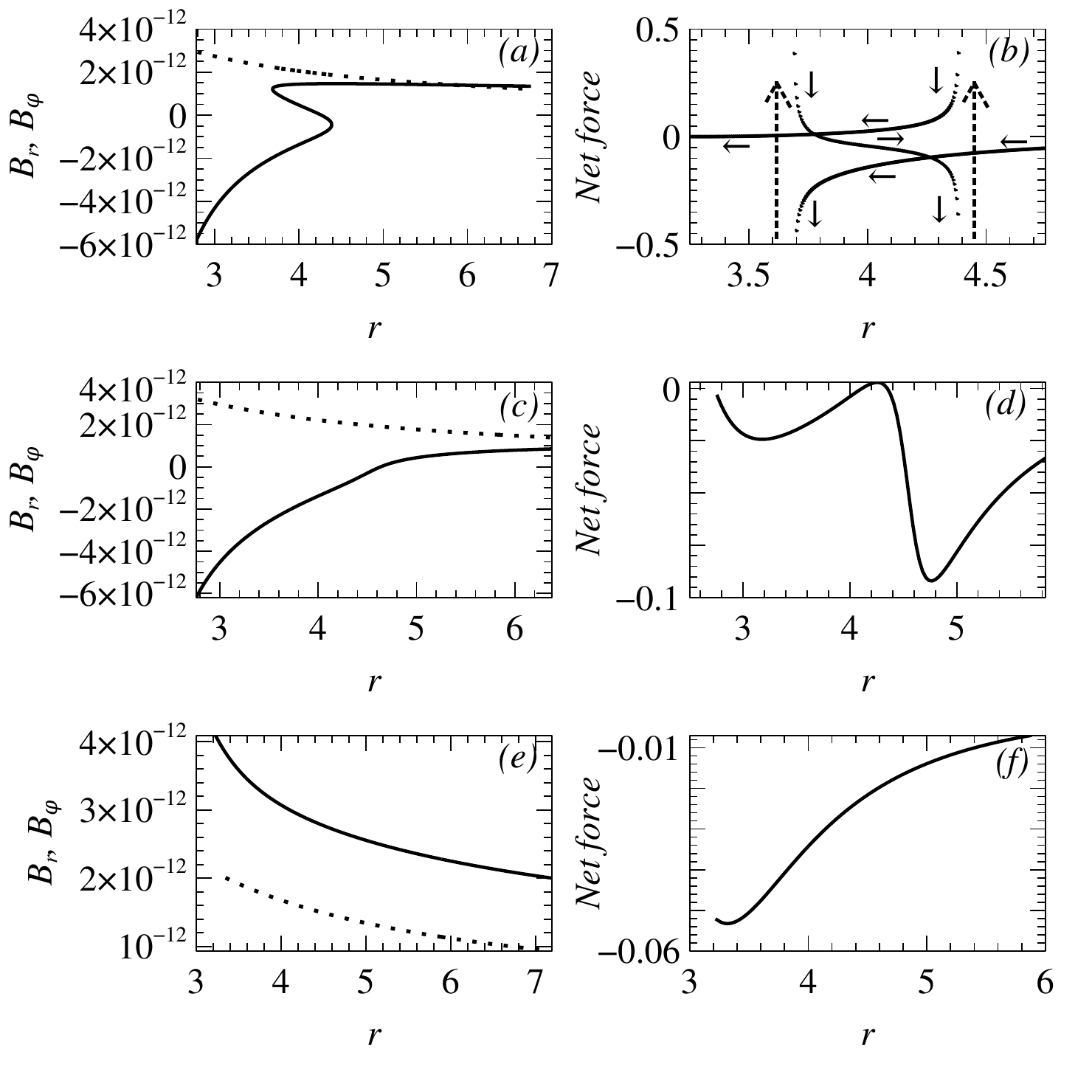}\caption{The radial (dotted lines) and azimuthal (solid lines) components of magnetic field to describe the origin of magnetic barrier for different relative field strength at the critical point as in Figure~\ref{fig:figure1}, such that at $r=r_{c}$, $(a)$ $B_{rc}=B_{\phi c}$, $(c)$ $B_{rc}=2B_{\phi c}$, $(e)$ $B_{rc}=B_{\phi c}/2$. The corresponding net forces are given in $(b)$, $(d)$ and $(f)$ respectively. In $(b)$ the long dotted arrows indicate the infinite discontinuous jump, whereas the small arrows indicate the direction of matter movement. The other parameters are same as in Figure~\ref{fig:figure2}.}
    \label{fig:figure3}
\end{figure*}

 Accretion disc can carry small as well as large scale magnetic fields, of course there is a certain upper limit to the amount of magnetic flux the disc around a black hole can sustain. However the strength of the magnetic field plays very crucial role in the dynamics of the accretion flow. The large scale field generally can not be produced in the disc. However, a seed magnetic field can be generated from zero initial field condition through, e.g., the Biermann battery mechanism \citep{Safarzadeh2017}, when there are non-aligned gradients in density and temperature profiles, automatically arising in the accretion disc structure. On the other hand, the externally generated field can be captured from the environment, say, interstellar medium and dragged inward by an accretion flow. This weak magnetic field can be dynamically dominant through flux freezing due to the inward advection of the magnetic field in this quasi-spherical accretion flow. This large amount of poloidal magnetic field cannot be escaped due to continued inward accretion pressure and also cannot be absorbed by the central black hole. In this situation, matter has to fight on its way to fall to the event horizon and facing this type of magnetic barrier. The origin of this magnetic barrier is described elaborately in Figures~\ref{fig:figure2} and \ref{fig:figure3}.

Figure~\ref{fig:figure2} shows the solutions for some important dynamical variables for three different relative field configurations: $B_{rc}=B_{\phi c}$ (for solid line), $B_{rc}=B_{\phi c}/2$ (for dashed line) and $B_{rc}=2B_{\phi c}$ (for dotted line) at $r_{c}=5.9$, where $\lambda_{c}=3.2$ is same for all the three cases. Note that here and subsequent cases discussed below, similar respective results are possible to obtain for highly relativistic ($\gamma\sim 4/3$) as well as less relativistic (with $\gamma\sim 1.4$ or so) flows with the slight readjustment of $r_c$ or $\lambda_c$.
 In Figure~\ref{fig:figure2}$(a)$ for the Mach number profile, the solid curve indicates that initially matter drags inward and $M$ increases gradually until it faces the magnetic barrier at $r\approx 3.7$. While it tries to go away from black hole, again faces other barrier at $r\approx 4.4$ and then slowly falls back onto the event horizon. The dashed line indicates that $M$ increases gradually as matter drags inward without facing any magnetic barrier but it is always in the sub-sonic region. The dotted one shows that even there is no magnetic barrier, magnetic field arrests the infalling matter and slows it down to the horizon. The corresponding specific angular momentum profile is shown in Figure~\ref{fig:figure2}$(b)$. Here, angular momentum transport is happening by large scale magnetic stress. The sound speed and the $Plasma-\beta$ are also shown in Figure~\ref{fig:figure2}$(c)$ and \ref{fig:figure2}$(d)$ respectively. The details of dynamics will be explained in the next section, however we try to address here the origin of magnetic barrier only. 

Figures~\ref{fig:figure3}$(a)$, \ref{fig:figure3}$(c)$ and \ref{fig:figure3}$(e)$ show the profiles for magnetic field components for three different conditions at the critical point and the corresponding net forces are given in Figures~\ref{fig:figure3}$(b)$, \ref{fig:figure3}$(d)$ and \ref{fig:figure3}$(f)$. Here, the net force indicates sum over all the forces as mentioned above and it will be negative when inward supporting gravity forces dominate. Symbolically, the net force is R.H.S. of equation as given below
\begin{eqnarray}
  v\frac{dv}{dr}=\frac{\lambda^{2}}{r^{3}}-\frac{1}{\rho}\frac{dp}{dr}-F-\frac{1}{4\pi \rho}\left(B_{\phi}\frac{dB_{\phi}}{dr}+\frac{B_{\phi}^{2}}{r}\right) \label{eq:netforce}.
\end{eqnarray}
Figure~\ref{fig:figure3}($a$) shows that the disc is poloidally dominated before flow faces barrier at $r\approx 3.7$ and the field components satisfy the barrier conditions as given in equation~(\ref{eq:barrierIa}). The corresponding net force is shown in Fig.~\ref{fig:figure3}$(b)$, initially which increases gradually to larger negative values as matter drags inward and jumps discontinuously at the barrier location at $r\approx 3.7$ from negative to positive direction indicating matter faces a negative impulse and tries to go back to infinity. The long dotted arrows indicate the infinite discontinuous jump and the small arrows indicate the direction of matter movement. Fig.~\ref{fig:figure3}($e$) shows that the disc is toroidally dominated and the net force is always negative shown in Fig.~\ref{fig:figure3}$(f)$. Hence the magnetic barrier does not appear and matter drags inward freely. In between these two cases, in Figures~\ref{fig:figure3}$(c)$ and \ref{fig:figure3}$(d)$, the barrier tries to appear in such a way that the poloidal and toroidal magnetic fields arrest the infalling mass and slows it towards the black hole. Hence, the accretion flow is decelerated near the black hole by large scale magnetic field.

What will be the fate of matter after knocking the barrier? As shown in Figures~\ref{fig:figure2} (solid line), \ref{fig:figure3}$(a)$ and \ref{fig:figure3}$(b)$, after knocking the barrier, matter will try to go away from the black hole. In this context, the behaviour of magnetic stress-tensors is very important, since the field lines already arrest the particles. The magnetic stress acts like a negative pressure along the field lines, as is in the case of a stretched elastic wire and this negative stress is known as `magnetic tension' (see \citealt{Spruit2013}). On the way going away from black hole, matter totally loses its angular momentum and the cumulative action of inward strong gravity and the magnetic tension along the field line controls the system. Hence, the matter is prevented from escaping due to the dominant nature of the net inward supporting forces. This is the origin of the second magnetic barrier as shown in Figures~\ref{fig:figure3}$(a)$ and \ref{fig:figure3}$(b)$ at $r\approx 4.4$.

\subsection{Disc dynamics}\label{sec:dynamics}

Now we concentrate on disc flow behaviours in details. Depending on the location of the critical points and the corresponding relative magnetic field strengths, the size of the disc with sub-Keplerian flow varies. The conditions for all different cases are supplied in Table~\ref{tab:table1}.
%----------------------------------------------------------------------------
\begin{table}
	\centering
	\caption{Conditions at the critical point and at outer boundary.}
	\label{tab:table1}
	\begin{tabular}{lcccr} % four columns, alignment for each
		\hline
		Figure & $r_{out}$ & $r_{c}$ & $\lambda_{c}$ & $B_{ic}$\\
		\hline
		\ref{fig:figure2} \& \ref{fig:figure3} & 6.74 & 5.87 & 3.2 & $B_{rc}=B_{\phi c}$\\
		\ref{fig:figure5} & 67 & 25.1 & 3.2 & $B_{rc}=B_{\phi c}$\\
		\ref{fig:figure7} & 220 & 41 & 3.2 & $B_{rc}=B_{\phi c}$\\
		\ref{fig:figure8} & 152.5 & 41 & 3.6 & $B_{rc}=B_{\phi c}$\\
		\ref{fig:figure2} \& \ref{fig:figure3} & 6.38 & 5.85 & 3.2 & $B_{rc}=2B_{\phi c}$\\
		\ref{fig:figure6} & 174 & 55 & 3.2 & $B_{rc}=2B_{\phi c}$\\
		\ref{fig:figure2} \& \ref{fig:figure3} & 7.19 & 5.9 & 3.2 & $B_{rc}= B_{\phi c}/2$\\
		\ref{fig:figure4} & 122 & 23 & 3.2 & $B_{rc}= B_{\phi c}/2$\\
		\hline
	\end{tabular}
\end{table}

\begin{figure}
\includegraphics[width=10cm]{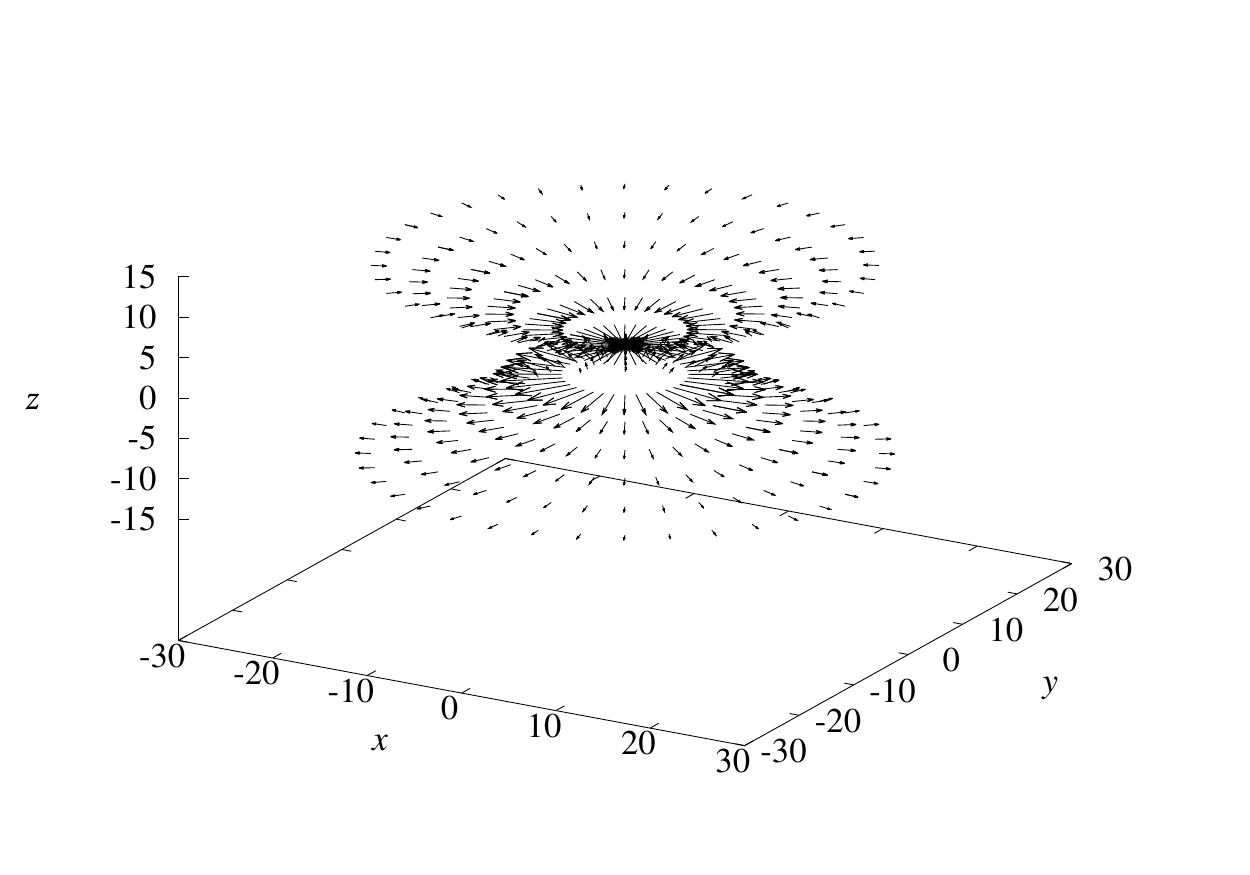}
\caption{The nature of magnetic field vectors in a typical model accretion flow 
around a black hole considered here. Vectors in both upper- and lower-half planes are considered together. }
    \label{fig:fig4}
\end{figure}

 Figure~\ref{fig:fig4} illustrates the nature of magnetic field lines in accretion flows around a black hole
considered here. Although our model explains behaviours of accretion flows on the upper-half of the disc (positive scale height only), 
from symmetry the lower half-plane can be visualized as well. Figure~\ref{fig:fig4} shows that the field lines direct towards the 
black hole in the upper half plane, whereas it is opposite in the lower-half plane. This is indeed expected in accordance with
equation (\ref{eq:nomonopole}) which, along with the solution for $B_{\phi}$, for the present model, gives rise to a split-monopole-like 
feature.

%----------------------------------------------------------------------

\begin{figure*}
\includegraphics[width=16cm]{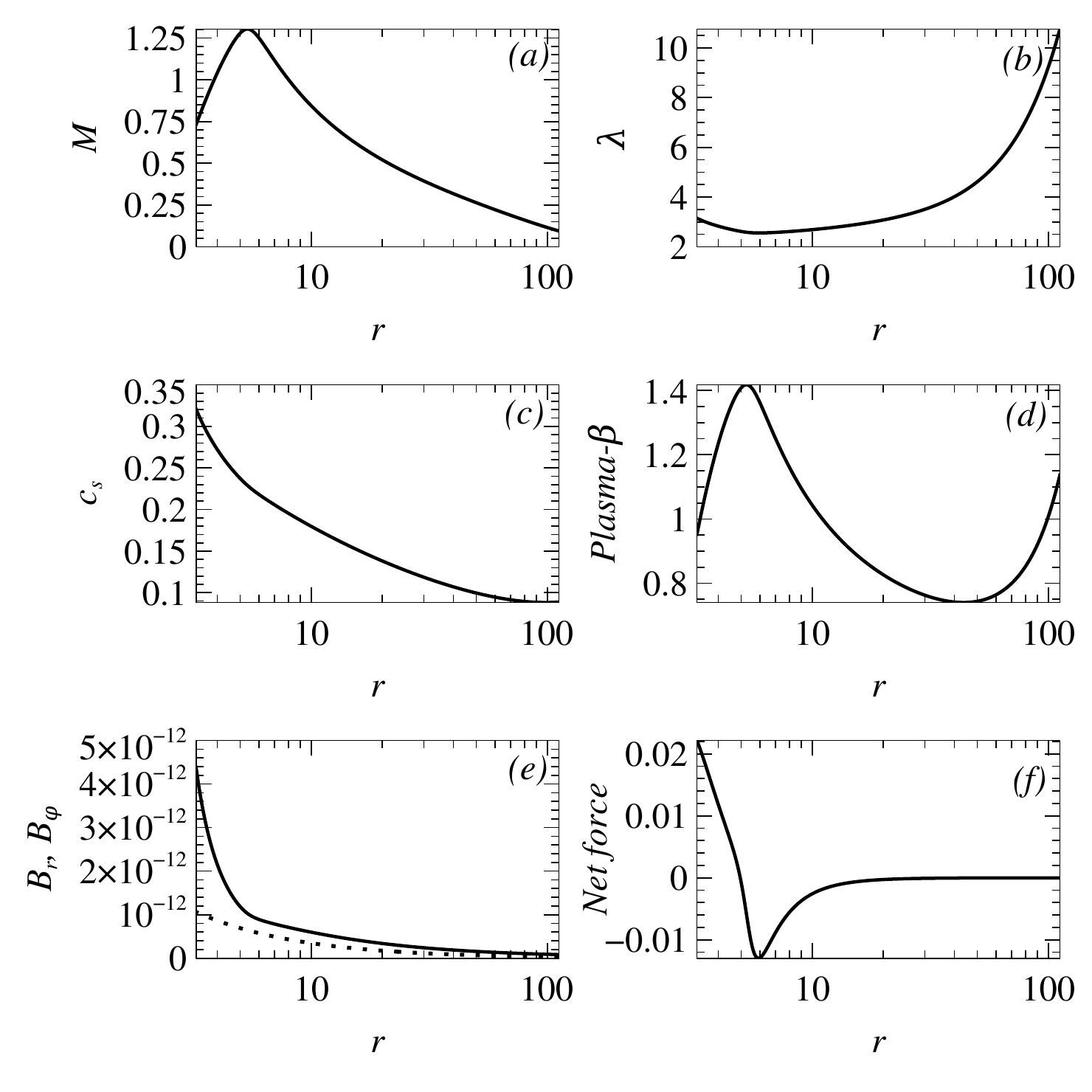}\caption{$(a)$ Mach number, $(b)$ angular momentum per unit mass, $(c)$ sound speed, $(d)$ plasma-$\beta$, $(e)$ the radial (dotted line) and azimuthal (solid line) components of magnetic field and (f) net force. Here, $r_{c}=23$, $r_{out}=122$, $\lambda_{c}=3.2$, $B_{rc}=B_{\phi c}/2$ and the other parameters are same as in Figure~\ref{fig:figure2}.}
    \label{fig:figure4}
\end{figure*}

The solution for some important dynamical and thermodynamical variables is shown in Figure~\ref{fig:figure4}, where $r_{c}=23$ and $r_{out}=122$. At $r_{c}$, the Mach number $M_{c}\approx 0.47$, i.e. $v_{c}\approx 0.47c_{sc}$. As matter advances towards event horizon, which is located at $r=2$ for non-rotating black hole, $M$ increases. This is quite common in accretion discs, but the interesting fact arises at $r\approx 5.3$ shown in Figure~\ref{fig:figure4}$(a)$, beyond which matter fails to accelerate and hence $M$ decreases. This is due to the dynamically dominant magnetic field. The strong magnetic field arrests the infalling matter and slows it down further towards black hole. The field components are shown in Figure~\ref{fig:figure4}$(e)$. Since the disc is toroidally dominated, infalling matter rotates more rather than its inward dragging. Figure~\ref{fig:figure4}$(f)$ indicates the corresponding net force given by equation~(\ref{eq:netforce}) and it will be negative when net inward force  dominates. Below $r\approx 5.3$, the net force indeed becomes positive, where the disc is already arrested by the field lines.

Figure~\ref{fig:figure4}$(b)$ shows that the outward transport of angular momentum is apparent in the presence of large scale magnetic stress, when at the outer boundary $r_{out} \approx 122$, $\lambda=\lambda_{K}$. However, in the inner region, where the disc is arrested mostly by toroidal magnetic field, $\lambda$ does not decrease and the infalling matter rotates very fast rather than accelerating inward and hence the velocity components $v\approx 0.23$ and $v_{\phi}=\lambda/r \approx 0.97$ near the event horizon. Figure~\ref{fig:figure4}$(c)$ shows that the sound speed increases monotonically towards the horizon and reaches a maximum of $0.32$, corresponding to a temperature $T\approx 10^{12}K$, as expected in advective accretion flow. Figure~\ref{fig:figure4}$(d)$ indicates the trend of $Plasma-\beta$ showing the magnetic pressure comparable to the fluid pressure and sometimes even more than that due to the presence of large scale strong magnetic field.  

\begin{figure*}
\includegraphics[width=16cm]{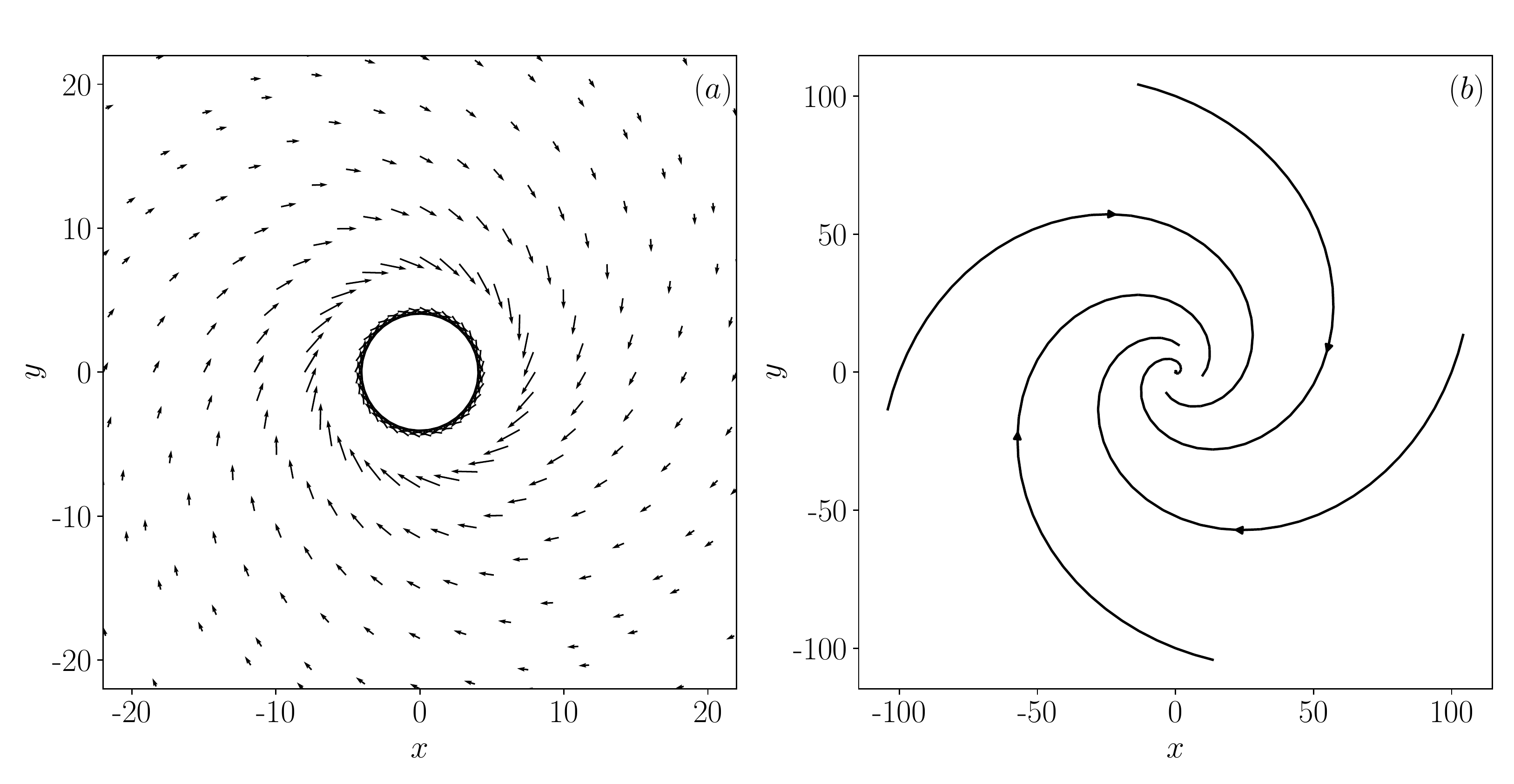}\caption{The nature of magnetic field $(a)$ vectors, and $(b)$ stream lines, in the 
$x-y$ plane of the accretion flow around a black hole for the case shown in Figure~\ref{fig:figure4}.}
    \label{fig:fig6}
\end{figure*}

\begin{figure}
\includegraphics[width=10cm]{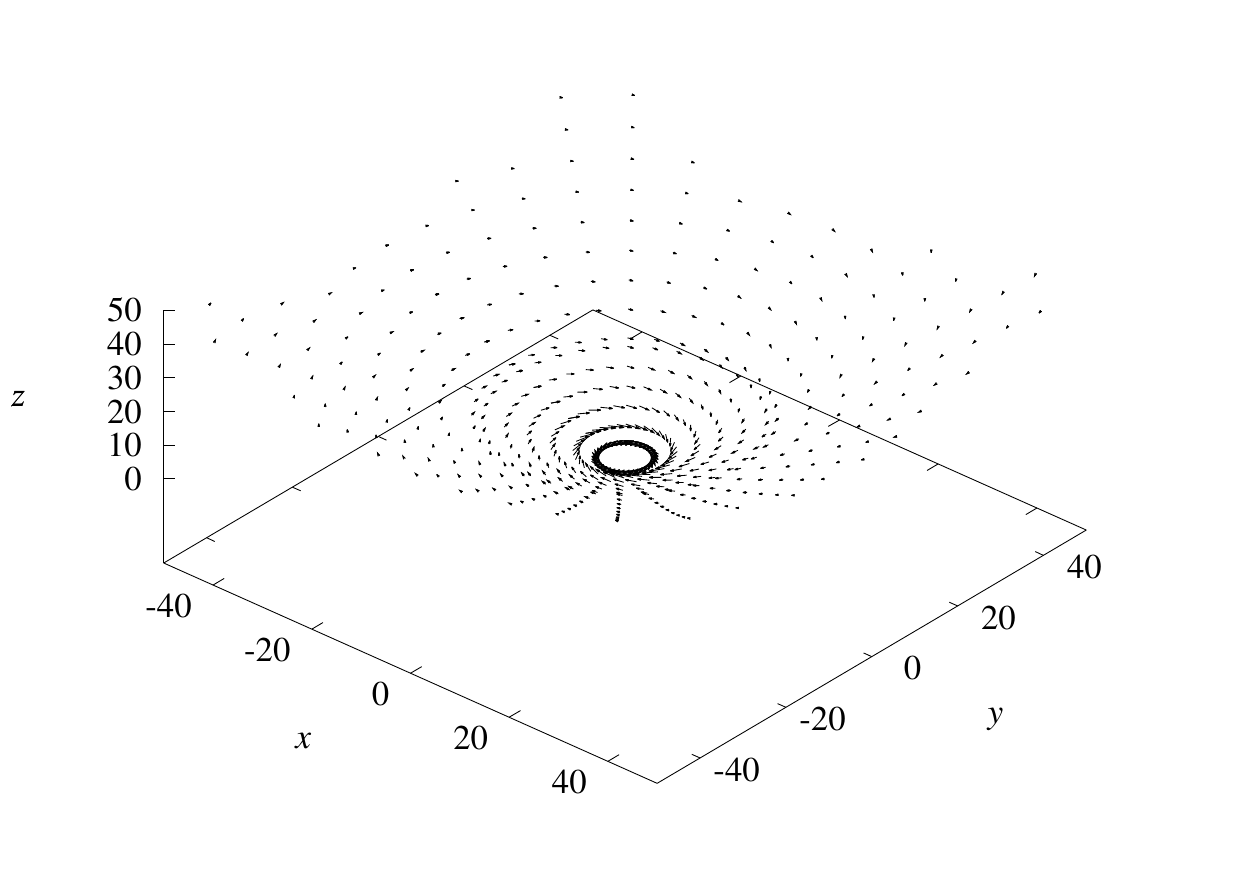}
\caption{Three-dimensional visualization of the magnetic field lines shown in
Figure~\ref{fig:fig6}.}
    \label{fig:fig7}
\end{figure}

 The nature of magnetic field vectors in the $x-y$ plane around a black hole for the above flow configuration is shown in 
Figure~\ref{fig:fig6} and also the corresponding three-dimensional visualization in the upper half plane of the disc is shown 
in Figure~\ref{fig:fig7}. Both the results are in accordance with Figure~\ref{fig:figure4} when matter is infalling throughout.

%------------------------------------------------------------------------

\begin{figure*}
 \includegraphics[width=16cm]{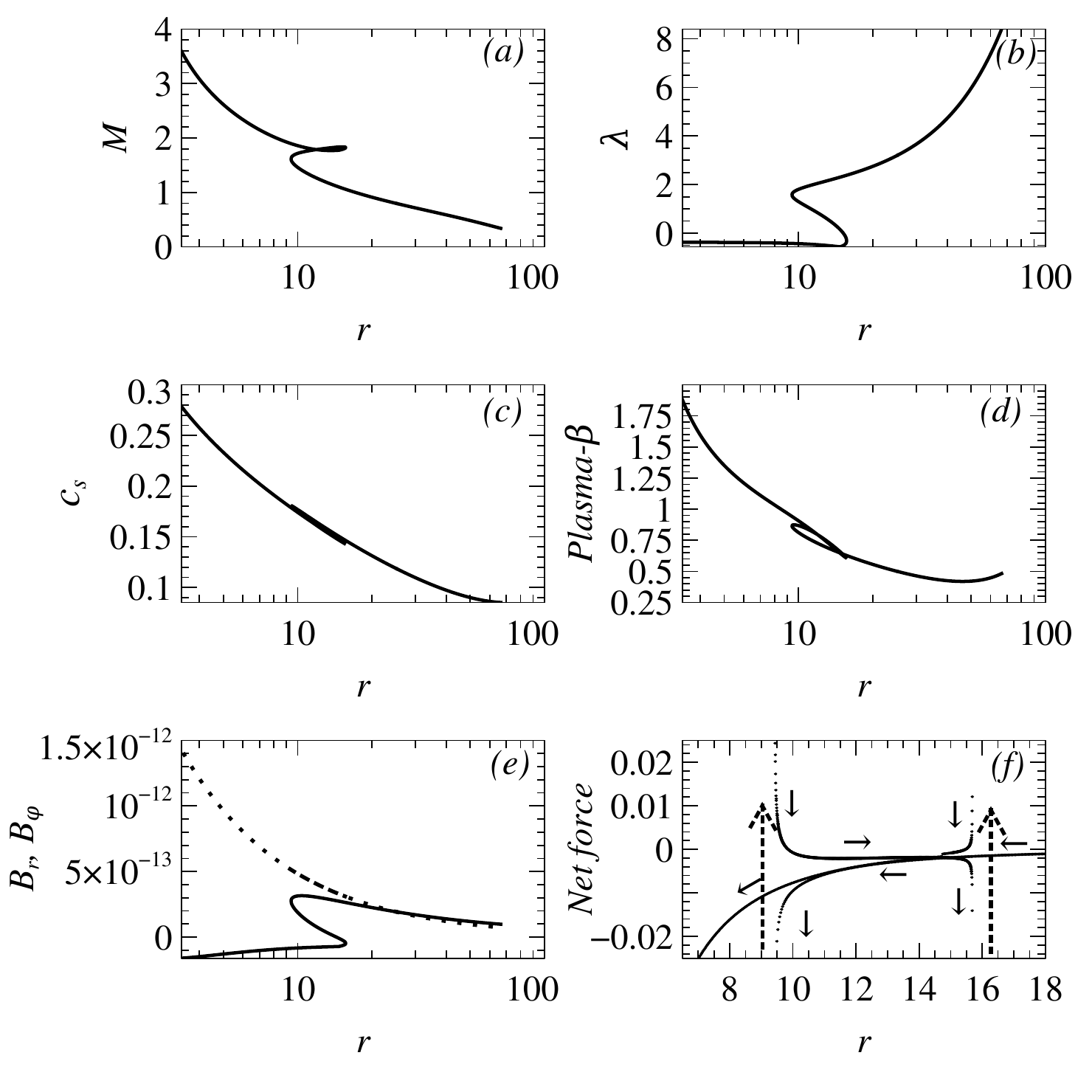}\caption{$(a)$ Mach number, $(b)$ angular  momentum per unit mass, $(c)$ sound speed, $(d)$ plasma-$\beta$, $(e)$ the radial   (dotted line) and azimuthal (solid line) components of magnetic field and $(f)$ net force. Here, $r_{c}=25.1$, $r_{out}=67$, $\lambda_{c}=3.2$, $B_{rc}=B_{\phi c}$ and the other parameters are same as in Figure~\ref{fig:figure2}.}
    \label{fig:figure5}
\end{figure*}

For different flow configurations, the solution for the same dynamical variables is shown in Figure~\ref{fig:figure5}. Here $r_{c}=25.1$. Figure~\ref{fig:figure5}$(a)$ shows Mach number profile and $M_{c}\approx 0.8$. Initially $M$ increases monotonically up to $r\approx 9.5$, where the first magnetic barrier appears due the accumulation of significant amount of poloidal magnetic field. After knocking the barrier matter tends to go away from the event horizon but on the way it again faces other barrier at $r\approx 15.6$ and hence the matter is prevented from escaping by the cumulative action of strong gravity and the magnetic tension along the field lines. The origin of these barriers are already explained in \S \ref{sec:barrier}. The small loop in the Mach number profile between these two barriers indicates the existence of `center' type or `O'-type critical point. Inside this region $M$ again increases and matter reaches the event horizon, where the radial velocity $v=1$. The corresponding net force field acting on the matter is shown in Figure~\ref{fig:figure5}$(f)$ and it will be negative if the net inward force dominates. Initially it increases negatively, indicating accretion phenomenon, but at the first barrier location $(r\approx 9.5)$ it shows infinitely discontinuous jump from negative to positive. Similar infinite discontinuous jump from negative to positive also happens at the outer barrier location $(r\approx 15.6)$. After that it again increases negatively, indicating accretion phenomenon. In Figure~\ref{fig:figure5}$(f)$, we particularly focus around the barrier regions.

Figure~\ref{fig:figure5}$(e)$ shows the field components, where the radial magnetic field follows $r^{-1}$ profile independently over the whole solution as given by no-monopole equation~(\ref{eq:nomonopole}), whereas the azimuthal part is coupled with other dynamical variables and can not be expressed in such a simple fashion. Near the first barrier location (at $r\approx 9.5$), the disc is poloidally dominated as discussed in the previous section. The azimuthal magnetic field profile along with the specific angular momentum decides how big the loop (due to presence of `O'-type critical point) in $M$-profile is, in between the two barriers. It can be more visualized in next two cases.

Figure~\ref{fig:figure5}$(b)$ shows that the outward transport of angular momentum occurs in the presence of large scale magnetic stress when the outer boundary corresponds to $\lambda=\lambda_{K}$ is at $r_{out}\approx 67$. Depending on the conditions at the critical point, the slope $\partial \lambda/ \partial r$ here is more compared to the case as shown in Figure~\ref{fig:figure4} and, hence, the steeper $\lambda$ profile helps the matter to lose the angular momentum faster, making the disc size smaller compared to the case as in Figure~\ref{fig:figure4}. Initially $\partial \lambda/ \partial r$ is positive, indicating matter is dragging inward. After knocking the first barrier at $r\approx 9.5$, matter tends to go away from the black hole with $\partial \lambda/ \partial r <0$. But on this way, matter totally loses its angular momentum and due to the dominant nature of the net inward forces, matter again faces other barrier at $r\approx 15.6$ and falls back to horizon. Figure~\ref{fig:figure5}$(c)$ shows that the sound speed reaches a maximum of $0.28$ at horizon, corresponding to a temperature $T\gtrsim 10^{11}K$. Figure~\ref{fig:figure5}$(d)$ shows the trend of $Plasma-\beta$ indicating the domination of magnetic pressure over fluid pressure almost throughout the flow.

%------------------------------------------------------------------------

\begin{figure*}
\includegraphics[width=16cm]{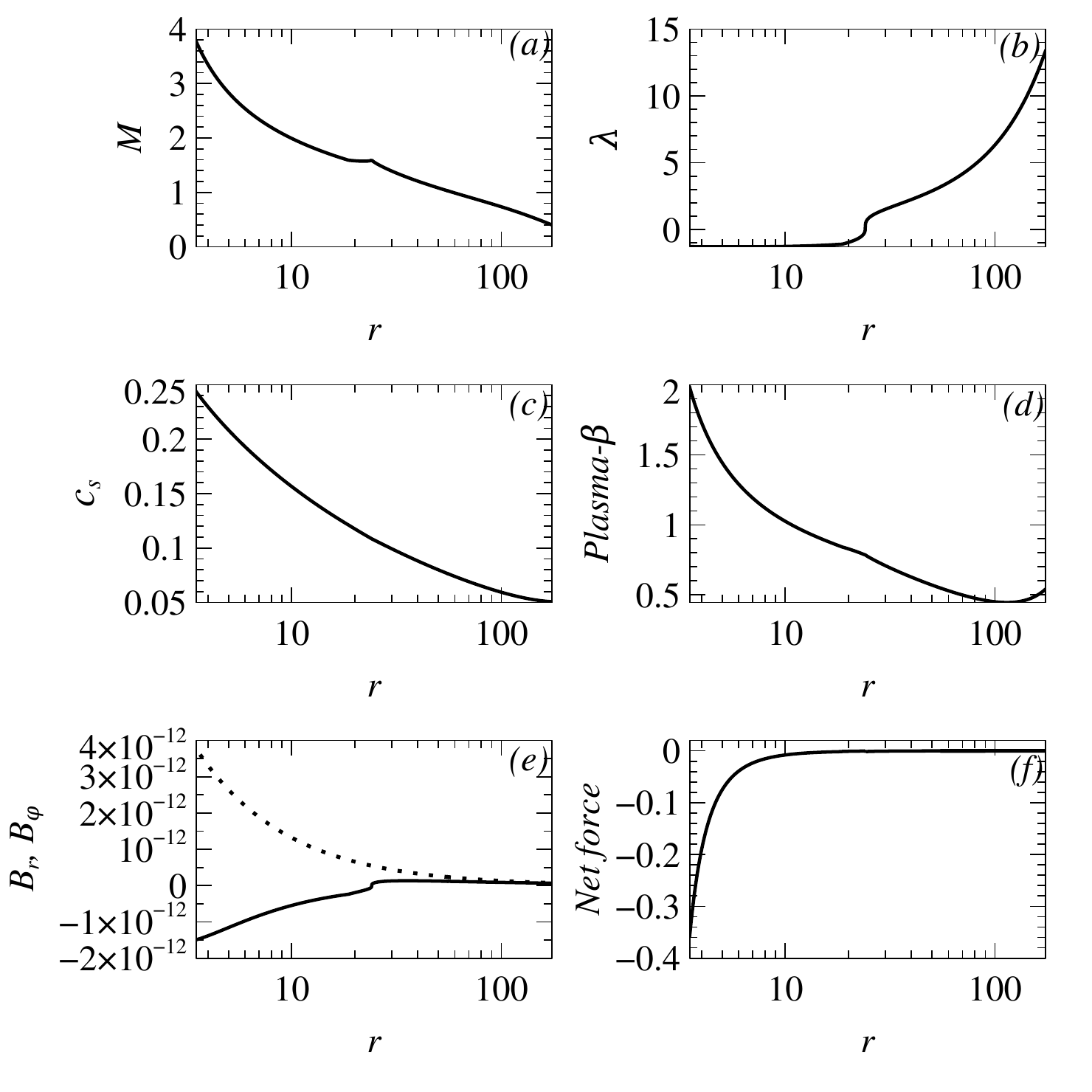}\caption{$(a)$ Mach number, $(b)$ angular momentum per unit mass, $(c)$ sound speed, $(d)$ plasma-$\beta$, $(e)$ the radial (dotted line) and azimuthal (solid line) components of magnetic field and $(f)$ net force. Here, $r_{c}=55$, $r_{out}=174$, $\lambda_{c}=3.2$, $B_{rc}=2B_{\phi c}$ and the other parameters are same as in Figure~\ref{fig:figure2}.}
    \label{fig:figure6}
\end{figure*}

Next we address disc dynamics for the case, where those two barriers merge to a single point in such a way that there will be no `O'-type critical point in between these barriers. This is shown in Figure~\ref{fig:figure6} when $r_{c}=55$, $B_{rc}=2B_{\phi c}$, $\lambda_{c} = 3.2$ and  $M_{c}=1.03$. Figure~\ref{fig:figure6}$(a)$ shows that the Mach number profile has a small kink at $r\approx 23$, which indicates the merging point of two barriers. Hence, $M$ increases monotonically as is in a accretion flow and it reaches a maximum at the event horizon, where the radial velocity becomes unity. Figure~\ref{fig:figure6}$(b)$ confirms the outward transport of $\lambda$ occurring by the large scale magnetic stress and the outer boundary corresponding to $\lambda=\lambda_{K}$ is at $r_{out} \approx 174$. Figure~\ref{fig:figure6}$(c)$ shows the sound speed basically carrying the information of the temperature of the disc. Figure~\ref{fig:figure6}$(d)$ and \ref{fig:figure6}$(e)$ show the $Plasma-\beta$ parameter and the magnetic field components respectively. The net force in Figure~\ref{fig:figure6}$(f)$ increases negatively as matter drags inward, towards the event horizon, as expected in an accretion flow.

%------------------------------------------------------------------------

\begin{figure*}
\includegraphics[width=16cm]{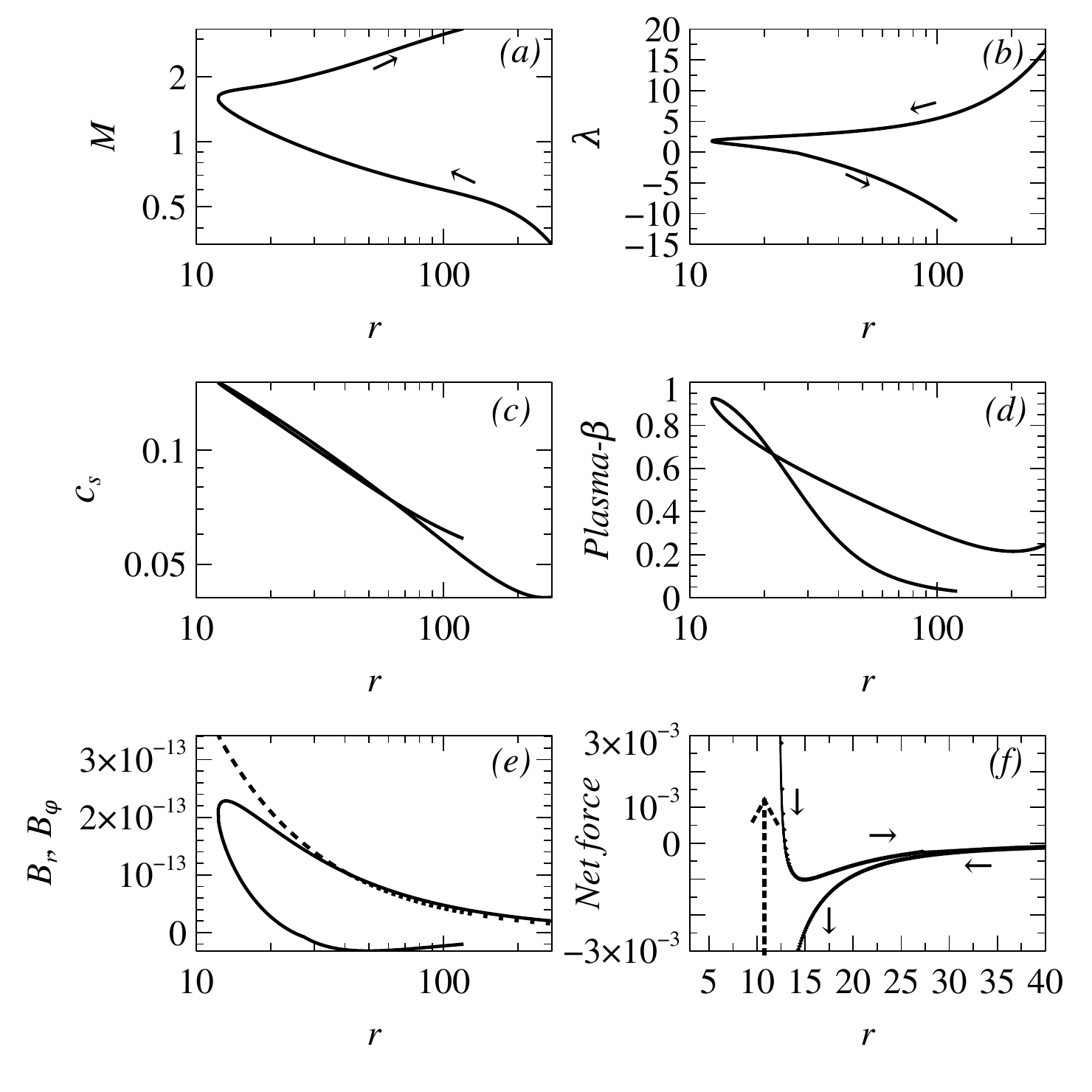}\caption{$(a)$ Mach number, $(b)$ angular momentum per unit mass, $(c)$ sound speed, $(d)$ plasma-$\beta$, $(e)$ the radial (dotted line) and azimuthal (solid line) components of magnetic field and $(f)$ net force. Here, $r_{c}=41$, $r_{out}=220$, $\lambda_{c}=3.2$, $B_{rc}=B_{\phi c}$ and the other parameters are same as in Figure~\ref{fig:figure2}.}
    \label{fig:figure7}
\end{figure*}

Figure~\ref{fig:figure7} shows a very important and unique solution. Here, the critical point is located at $r_{c}=41$ and at this point $B_{rc}=B_{\phi c}$ and $\lambda_{c}=3.2$. Figure~\ref{fig:figure7}$(a)$ shows that $M$ increases continuously as matter drags inward and due to accumulation of large amount of poloidal magnetic flux, matter faces the magnetic barrier at $r\approx 12.5$. After knocking the barrier, matter again goes back to infinity. This type of profile must have very significant contribution to outflow/jet. In a more realistic
three-dimensional model, matter would go vertically after knocking the barrier revealing outflow. The angular momentum profile in Figure~\ref{fig:figure7}$(b)$ shows that initially from the outer boundary corresponding to $\lambda=\lambda_{K}$ at $r_{out}=220$, $\lambda$ decreases as matter drags inward and after knocking the barrier it increases gradually for larger orbits and reaches the Keplerian limit again at $r\approx 110$. The negative sign just signifies that the matter is rotating in opposite direction after facing the barrier. The sound speed profile in Figure~\ref{fig:figure7}$(c)$ shows that initially it increases up to $0.15$ at the barrier location and then decreases monotonically as matter goes away from the black hole, which is expected due to lowering the potential energy. In Figure~\ref{fig:figure7}$(d)$, the $Plasma-\beta$ is less than unity, which indicates that the magnetic pressure is dominating over normal fluid pressure. The origin of this type of profile is hidden in magnetic field strength profile, given in Figure~\ref{fig:figure7}$(e)$. As usual the matter faces the magnetic barrier at $r\approx 12.5$ due to dominant behaviour of poloidal magnetic field in the inner region. After knocking the barrier matter goes far away from the black hole. Since, $B_{\phi}$ and as well as $\partial B_{\phi}/ \partial r$ become very weak gradually, magnetic tension is not large enough to prevent the matter being escaping. The net force in Figure~\ref{fig:figure7}$(f)$ also shows that initially it increases gradually in negative direction as matter drags inward and jumps discontinuously at the barrier location from negative to positive. The arrows indicate the direction of matter.

\begin{figure*}
\includegraphics[width=16cm]{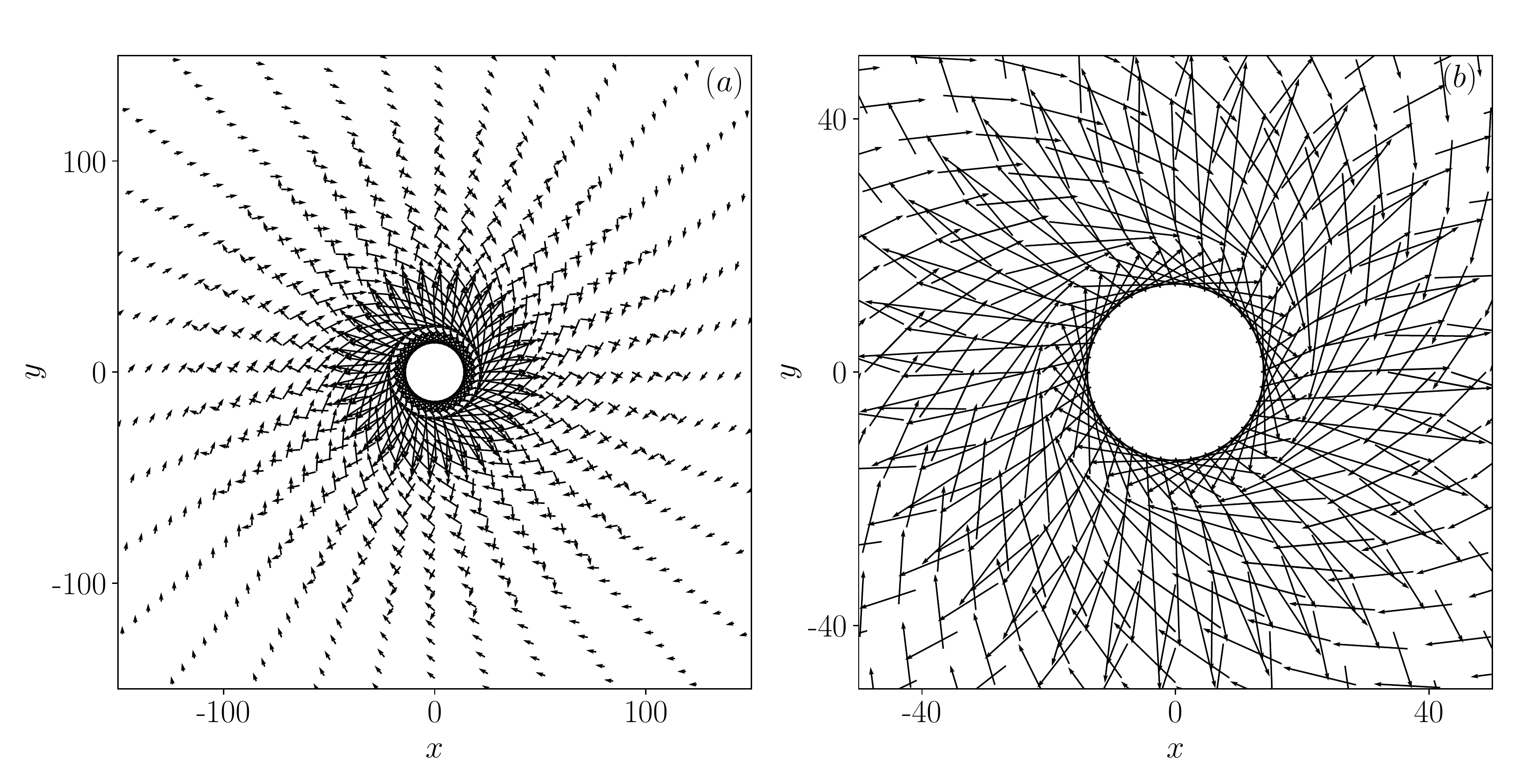}\caption{ The nature of $(a)$ magnetic field vectors, and $(b)$ magnetic field vectors near the 
barrier location, in the $x-y$ plane of the accretion flow around a black hole for the case shown in Figure~\ref{fig:figure7}.}
    \label{fig:fig11}
\end{figure*}

\begin{figure}
\includegraphics[width=10cm]{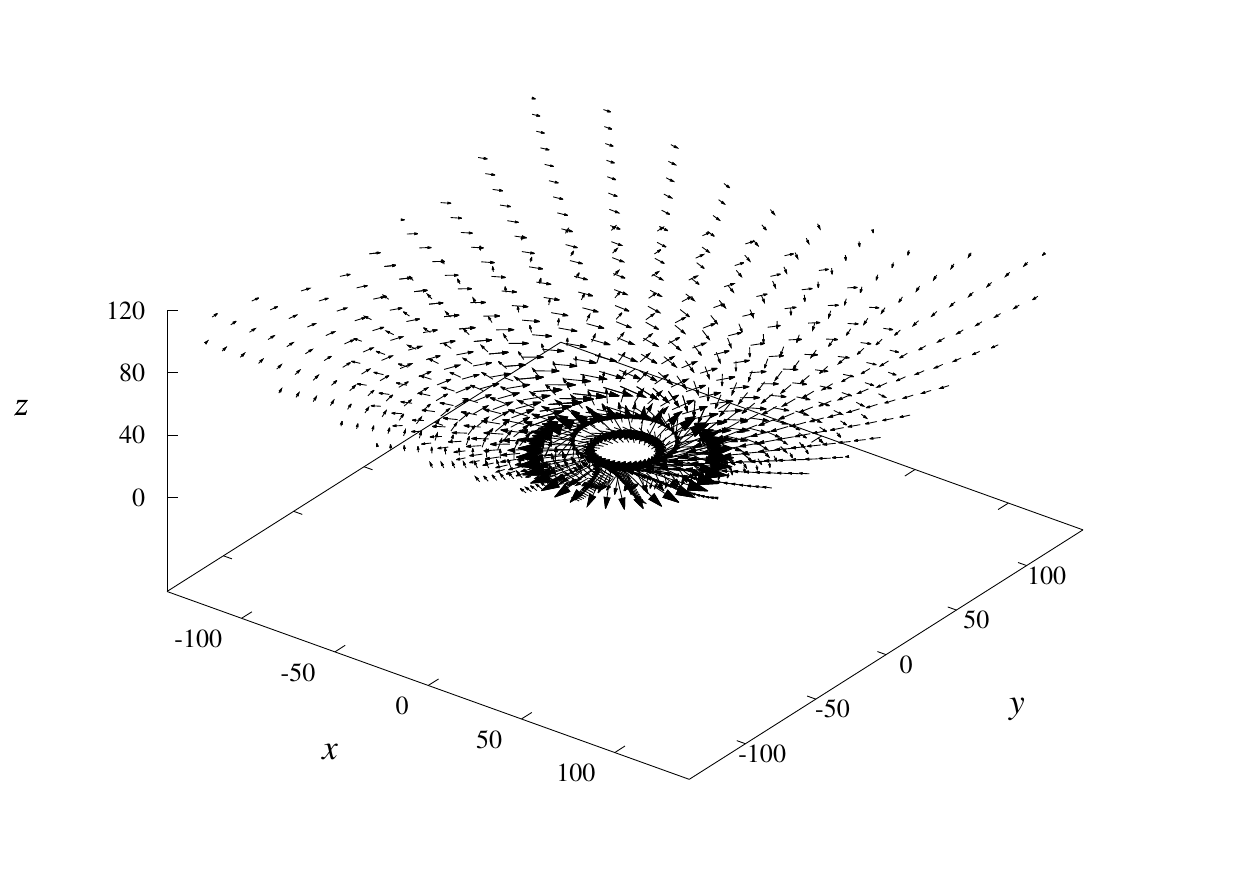}
\caption{Three-dimensional visualization of the magnetic field lines shown in 
Figure~\ref{fig:fig11}.}
    \label{fig:fig12}
\end{figure}

 The nature of magnetic field vectors in the $x-y$ plane for this unique flow is shown in
Figure~\ref{fig:fig11} and also the corresponding three-dimensional visualization in the upper half plane of the disc is shown
in Figure~\ref{fig:fig12}. Both the figures depict the absence of any magnetic vector in the inner flow region in accordance 
with Figure \ref{fig:figure7}. It is important to note that in a region $220\ge r\ge 12.5$, when the matter is falling in, 
the field vectors are in the inward direction. On the other hand, in $12.5\le r\le 110$, where the matter is flowing out, 
the field vectors are in the outward direction. Hence, there is a zone in Figures \ref{fig:fig11} and \ref{fig:fig12} where
the field lines of either directions appear simultaneously.

%------------------------------------------------------------------------

\begin{figure*}
\includegraphics[width=16cm]{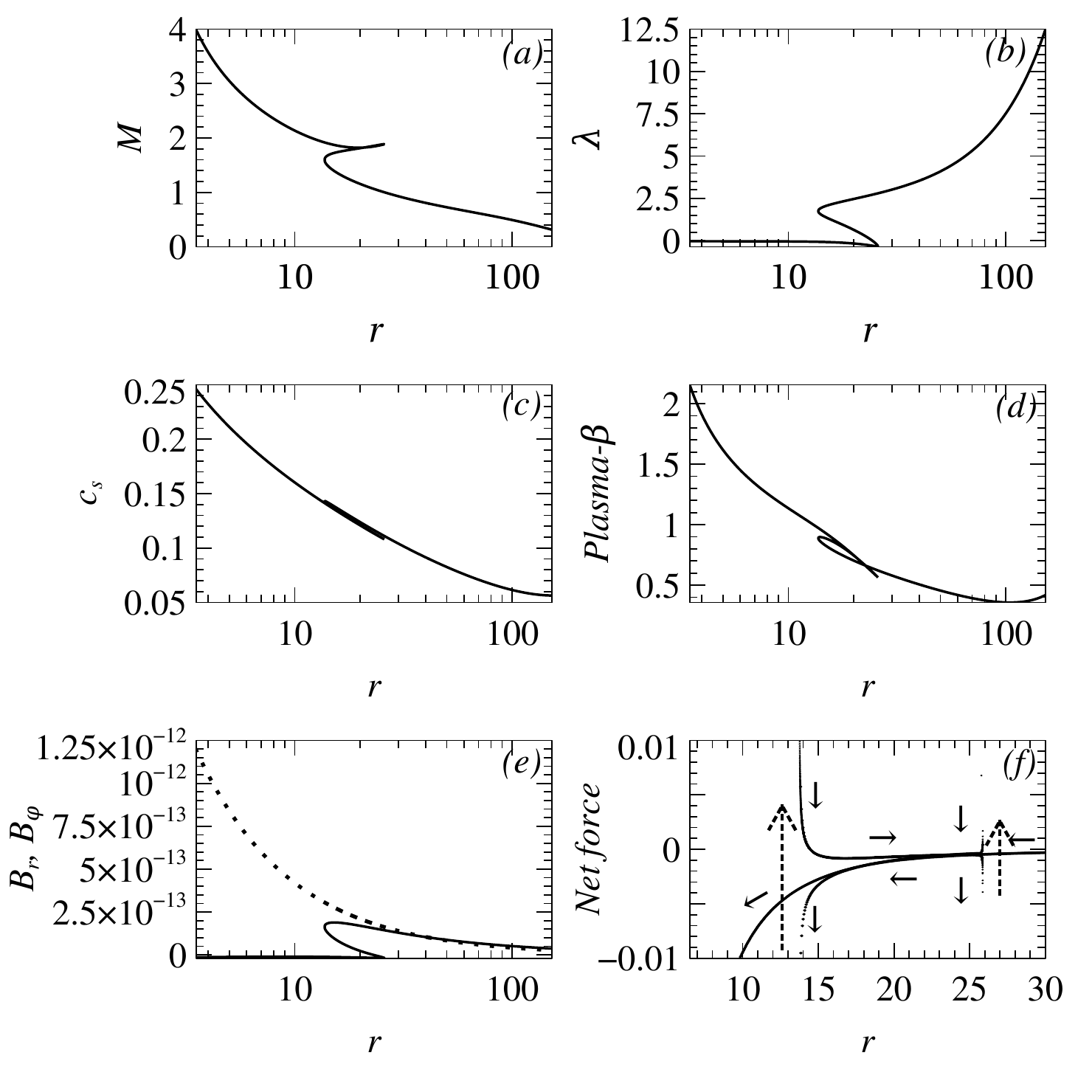}\caption{$(a)$ Mach number, $(b)$ angular momentum per unit mass, $(c)$ sound speed, $(d)$ plasma-$\beta$, $(e)$ the radial (dotted line) and azimuthal (solid line) components of magnetic field and $(f)$ net force. Here, $r_{c}=41$, $r_{out}=152.5$, $\lambda_{c}=3.6$, $B_{rc}=B_{\phi c}$ and the other parameters are same as in Figure~\ref{fig:figure2}.}
    \label{fig:figure8}
\end{figure*}

The solutions shown in Figure ~\ref{fig:figure8} are obtained with the same condition as of Figure~\ref{fig:figure7} except 
$\lambda_{c}$, which is 3.6 instead of 3.2. The importance of this figure is that it shows how the second barrier can arise with the presence of `O'-type critical point in between two barriers based on the relative dependence of specific angular momentum and magnetic field strength at the critical point. The azimuthal magnetic field profile strongly links with $\lambda$ according to equation~(\ref{eq:angmomentum}). Hence, the large angular momentum at critical point makes the slope $\partial \lambda / \partial r$ larger compared to that in Figure~\ref{fig:figure7}, which not only makes the disc size smaller but also makes the slope $\partial B_{\phi} / \partial r$ larger even after knocking the first barrier. Therefore, net inward force dominates due to magnetic contribution as given in equation~(\ref{eq:netforce}) and the second barrier appears. In other word, in this way these quantities just determine how big or small of the loop size in between these two barriers is in $M$ profile around the `O'-type critical point.

\begin{figure}
\includegraphics[width=\columnwidth]{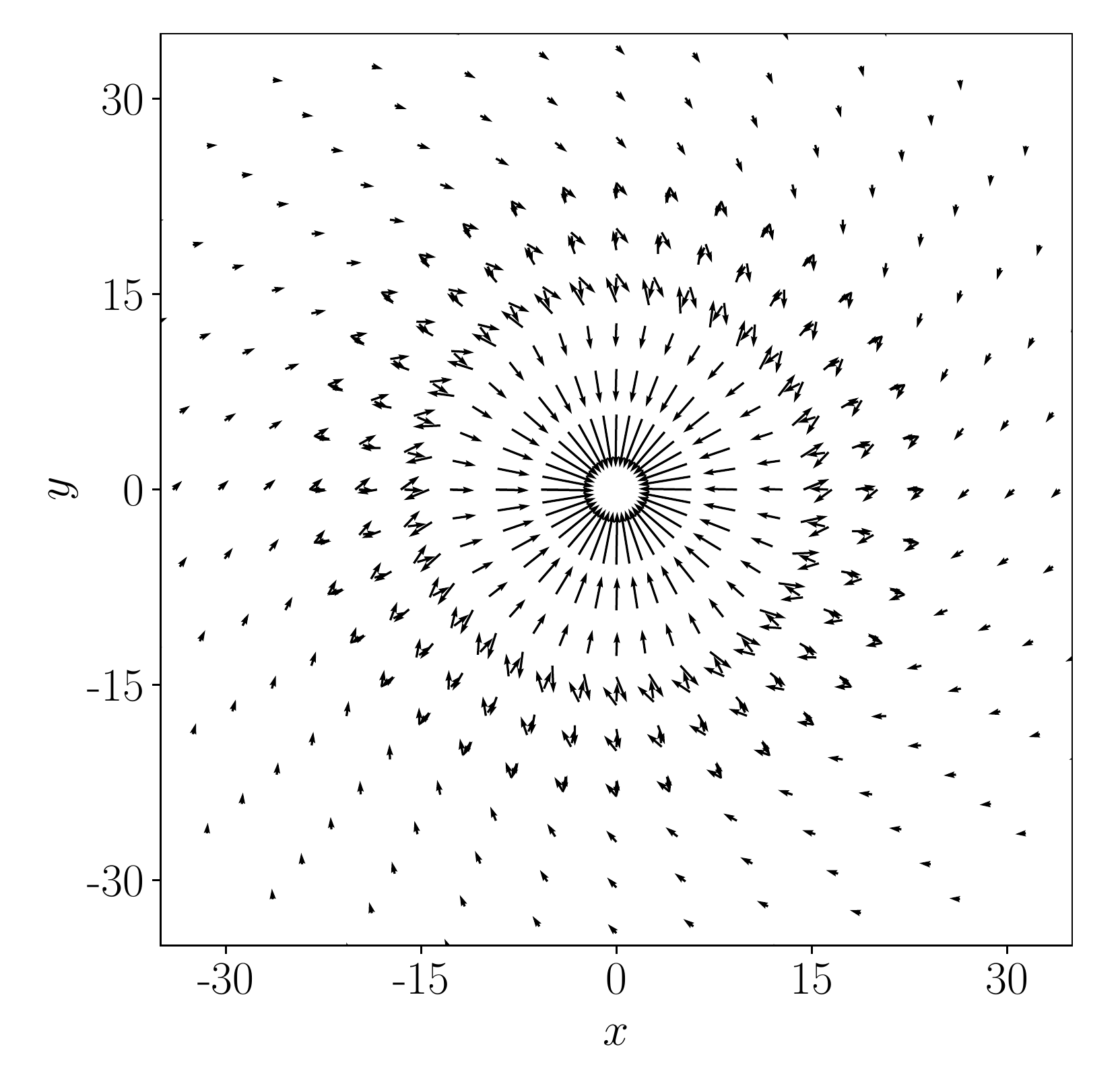}\caption{The nature of magnetic field vectors in the $x-y$ plane of the accretion 
flow around a black hole for the case shown in Figure~\ref{fig:figure8}.}
    \label{fig:fig14}
\end{figure}

\begin{figure}
\includegraphics[width=\columnwidth]{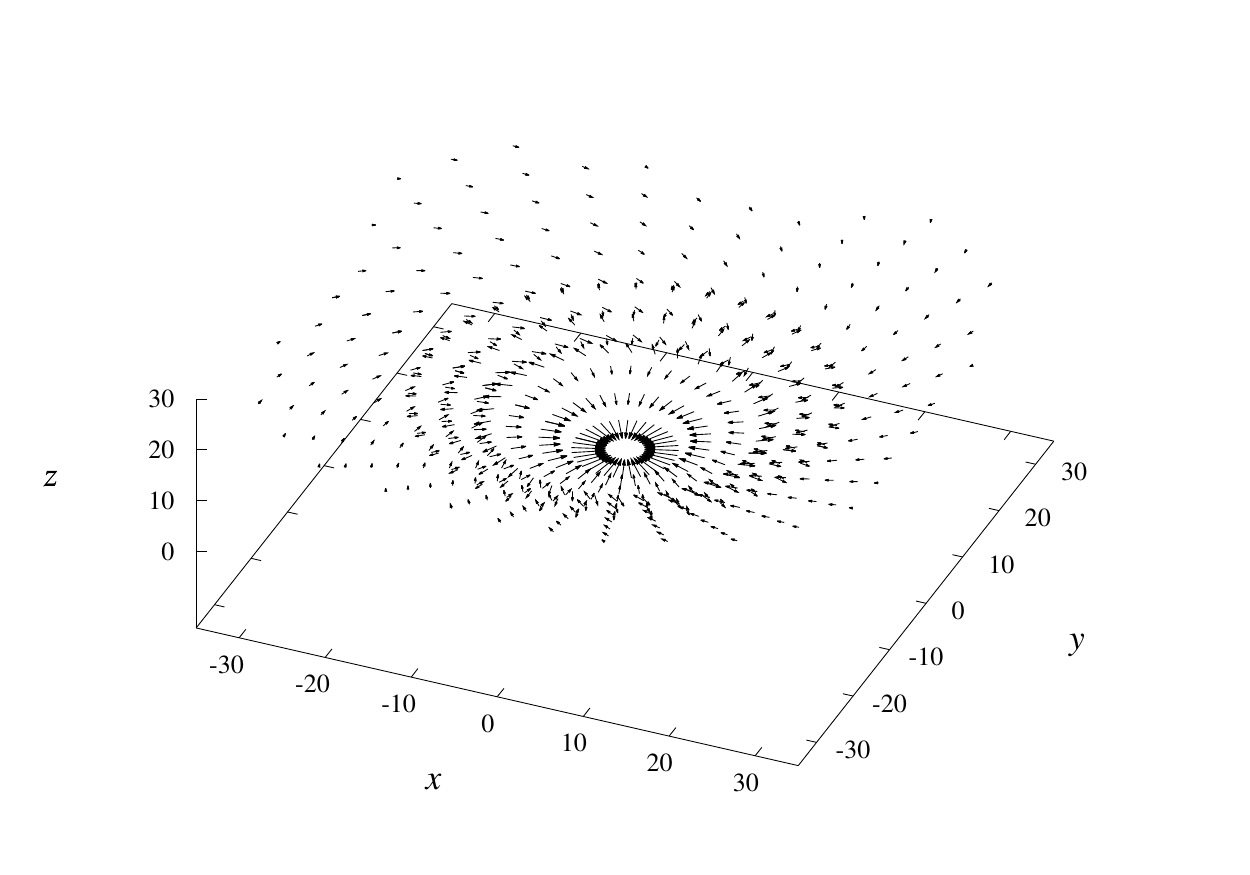}\caption{Three dimensional visualization of the magnetic field lines shown in 
Figure~\ref{fig:fig14}.}
    \label{fig:fig15}
\end{figure}

 The nature of magnetic field vectors in the $x-y$ plane for this flow is shown in
Figure~\ref{fig:fig14} and also the corresponding three-dimensional visualization in the upper half plane of the disc is shown
in Figure~\ref{fig:fig15}. Both the figures depict a zone where
the field lines of either directions appear simultaneously. This corresponds to the region of loop around `O'-type critical point.

%------------------------------------------------------------------------

\section{DISCUSSION AND CONCLUSIONS}\label{sec:discussion}

We have explored the effects of large scale strong magnetic field in the advective accretion flow in order to transport angular momentum, as well as the origin of different magnetic barriers in the MAD regime. Here, the radial accretion velocity is typically high compared to standard thin disc model and is proportional to magnetic tension. This is because the radial velocity is determined by how fast the  magnetic tension can transfer angular momentum outwards. The specific angular momentum of the flow is much smaller than the local Keplerian value and the outer boundary of this model corresponds to $\lambda=\lambda_{K}$, the beginning of the sub-Keplerian flow far away from the black hole. Hence, the disc size principally depends on magnetic field strength. 
 We have demonstrated the possible formation of four distinct flow classes: (1) no barrier and matter 
reaches the black hole, (2) a barrier stops the infall and the matter goes back completely, (3) infalling 
matter faces two barriers but eventually reaches the black hole event horizon, and (4) matter decelerates 
near the event horizon and eventually falls into the black hole.

 The presence of magnetic field plays very crucial role in the dynamics of the accretion flow. The accretion disc can carry small as well as large scale magnetic field. However, there is an upper limit to the amount of magnetic field, the disc around a black hole can sustain in steady-state and it is achieved when an accretion flow reaches the MAD state. Generally the large scale magnetic field can not be produced in the disc, rather it can be captured from the environment, say, interstellar medium and dragged inward by the accretion flow. This magnetic field can be dynamically dominant near the event horizon through flux freezing due to the inward advection of the magnetic flux in this quasi-spherical accretion flow. The accumulated poloidal field can not be absorbed by the black hole and also can not be escaped due to continued inward accretion pressure. At this circumstance, matter faces the magnetic barrier and again goes back to infinity. It exhibits two Mach numbers at same radius, one corresponds to infalling matter whereas the other one for outgoing matter in this quasi-spherical accretion flow. These types of profile are expected to play a very crucial role in the generation of various kinds of outflow. More generally this may be the building block to produce jet/outflow. Further, the Bernoulli parameter $b$ (given in APPENDIX~\ref{sec:appendix A}) is positive here. It also implies that highly magnetized advective accretion flow provides a generic explanation of unbound matter and hence outflows and jets.

 This is not the whole story! Since in MAD state, the magnetic field strength is large enough, it can play in its own way. After knocking the barrier when matter tries to go away from the black hole, it totally loses its angular momentum. At the same time, the cumulative action of inward strong gravity and the magnetic tension along the field line could be large enough to prevent the matter being escaped as well. Hence, matter might face a second magnetic barrier and fall back to the black hole. This can be a possible explanation of the episodic jet phenomena, where the magnetic field can lock the matter in between these two barriers depending on the relative dependence between the specific angular momentum and the magnetic field strength.

 Is there any observational evidence of such strong magnetic field as discussed here? Here, we find from the magnetic field profile, the field strength in the inner region of the accretion disc is of the order of $10^{7}-10^{8}\, G$ for stellar mass black holes. Interestingly, these field values tally almost nearly with observation, based on a model relating the observed kinetic power of relativistic jet to the magnetic field of the accretion discs \citep{Garofalo2010,Piotrovich2014}.

 The sound speed in this magnetized advective flow is much higher than that in Shakura-Sunyaev disc and the ion temperature of the accreting  gas is nearly virial, order of $10^{12}$ K. This is expected since there are no sufficient cooling.
 
 In this present context, for simplicity, we have assumed that the flow to be vertically averaged without allowing any vertical component of the flow, but considering the maximum upper limit of the magnetic field strength in order to achieve the MAD regime. Our next move will be to investigate the coupled disc-outflow system more self-consistently by including the vertical components of the flow variables in this strong large scale magnetic field regime.

\section*{ACKNOWLEDGEMENTS}

The work was partly supported by the project funded by ISRO with research Grant No.
ISTC/PPH/BMP/0362.

%%%%%%%%%%%%%%%%%%%%%%%%%%%%%%%%%%%%%%%%%%%%%%%%%%

%%%%%%%%%%%%%%%%%%%% REFERENCES %%%%%%%%%%%%%%%%%%

% The best way to enter references is to use BibTeX:

%\bibliographystyle{mnras}
%\bibliography{example} % if your bibtex file is called example.bib

% Alternatively you could enter them by hand, like this:
% This method is tedious and prone to error if you have lots of references

%%%%%%%%%%%%%%%%%%%%%%%%%%%%%%%%%%%%%%%%%%%%%%%%%%

%%%%%%%%%%%%%%%%% APPENDICES %%%%%%%%%%%%%%%%%%%%%

\appendix

\section{The Bernoulli parameter}\label{sec:appendix A}

To obtain the Bernoulli parameter, we have to integrate the radial momentum balance equation~(\ref{eq:radmomentum}). In this equation, this following terms can be simplified using equations~(\ref{eq:angmomentum}) and (\ref{eq:induction2}) and in the absence of viscosity, as
\begin{eqnarray*}
\frac{B_{\phi}}{4\pi r \rho}\frac{d}{dr}(rB_{\phi})-\frac{\lambda^{2}}{r^{3}} = \frac{vB_{\phi}}{rB_{r}}\frac{d\lambda}{dr}-\frac{\lambda^{2}}{r^{3}}\\ = \frac{d}{dr}\left(\frac{\lambda vB_{\phi}}{rB_{r}}\right)-\lambda \frac{d}{dr}\left(\frac{vB_{\phi}}{rB_{r}}\right)-\frac{\lambda^{2}}{r^{3}}\\ = \frac{d}{dr}\left(\frac{\lambda vB_{\phi}}{rB_{r}}\right)-\lambda \frac{d}{dr}\left(\frac{\frac{\lambda B_{r}}{r}}{rB_{r}}\right)-\frac{\lambda^{2}}{r^{3}}\\=\frac{d}{dr}\left(\frac{\lambda vB_{\phi}}{rB_{r}}\right)-\lambda \frac{d}{dr}\left(\frac{\lambda}{r^{2}}\right)-\frac{\lambda^{2}}{r^{3}}\\=\frac{d}{dr}\left(\frac{\lambda vB_{\phi}}{rB_{r}}\right)-\frac{\lambda}{r^{2}}\frac{d\lambda}{dr}+\frac{\lambda^{2}}{r^{3}}\\=\frac{d}{dr}\left(\frac{\lambda vB_{\phi}}{rB_{r}}\right)-\frac{d}{dr}\left(\frac{\lambda^{2}}{2r^{2}}\right)
\end{eqnarray*}
Hence, the Bernoulli parameter $(b)$ can be written as 
\begin{equation}
b=\frac{1}{2}v^{2}+\frac{\gamma}{\gamma -1} \frac{p}{\rho}+\int F dr+\frac{\lambda v B_{\phi}}{r B_{r}}-\frac{\lambda^{2}}{2r^{2}}.
\end{equation}
Now, integrating equation~(\ref{eq:angmomentum}) without viscosity, we obtain
\begin{equation*}
4\pi r\rho h v(\lambda-\lambda_{in})=r^{2}hB_{r}B_{\phi}.
\end{equation*}
Assuming $\lambda_{in}=0$ for a non-rotating black hole, the Bernoulli parameter becomes
\begin{equation}
b=\frac{1}{2}v^{2}+\frac{\lambda^{2}}{2r^{2}}+\frac{\gamma}{\gamma -1} \frac{p}{\rho}+\int F dr+\frac{B_{\phi}^{2}}{4\pi \rho}-\frac{B_{r}B_{\phi}\lambda}{4\pi r\rho v}.
\end{equation}
%%%%%%%%%%%%%%%%%%%%%%%%%%%%%%%%%%%%%%%%%%%%%%%%%%

% Don't change these lines
\bsp	% typesetting comment
\label{lastpage}
\end{document}